\documentstyle[11pt,psfig,twoside]{article}

\begin{document}

\newcommand{\dd}{\,{\rm d}}
\newcommand{\ie}{{\it i.e.},\,}
\newcommand{\etal}{{\it et al.\ }}
\newcommand{\eg}{{\it e.g.},\,}
\newcommand{\cf}{{\it cf.\ }}
\newcommand{\vs}{{\it vs.\ }}
\newcommand{\zdot}{\makebox[0pt][l]{.}}
\newcommand{\up}[1]{\ifmmode^{\rm #1}\else$^{\rm #1}$\fi}
\newcommand{\dn}[1]{\ifmmode_{\rm #1}\else$_{\rm #1}$\fi}
\newcommand{\upd}{\up{d}}
\newcommand{\uph}{\up{h}}
\newcommand{\upm}{\up{m}}
\newcommand{\ups}{\up{s}}
\newcommand{\arcd}{\ifmmode^{\circ}\else$^{\circ}$\fi}
\newcommand{\arcm}{\ifmmode{'}\else$'$\fi}
\newcommand{\arcs}{\ifmmode{''}\else$''$\fi}
\newcommand{\MS}{{\rm M}\ifmmode_{\odot}\else$_{\odot}$\fi}
\newcommand{\RS}{{\rm R}\ifmmode_{\odot}\else$_{\odot}$\fi}
\newcommand{\LS}{{\rm L}\ifmmode_{\odot}\else$_{\odot}$\fi}

\newcommand{\Abstract}[2]{{\footnotesize\begin{center}ABSTRACT\end{center}
\vspace{1mm}\par#1\par
\noindent
{~}{\it #2}}}

\newcommand{\TabCap}[2]{\begin{center}\parbox[t]{#1}{\begin{center}
  \small {\spaceskip 2pt plus 1pt minus 1pt T a b l e}
  \refstepcounter{table}\thetable \\[2mm]
  \footnotesize #2 \end{center}}\end{center}}

\newcommand{\TableSep}[2]{\begin{table}[p]\vspace{#1}
\TabCap{#2}\end{table}}

\newcommand{\FigCap}[1]{\footnotesize\par\noindent Fig.\  %
  \refstepcounter{figure}\thefigure. #1\par}

\newcommand{\TableFont}{\footnotesize}
\newcommand{\TableFontIt}{\ttit}
\newcommand{\SetTableFont}[1]{\renewcommand{\TableFont}{#1}}

\newcommand{\MakeTable}[4]{\begin{table}[htb]\TabCap{#2}{#3}
  \begin{center} \TableFont \begin{tabular}{#1} #4 
  \end{tabular}\end{center}\end{table}}

\newcommand{\MakeTableSep}[4]{\begin{table}[p]\TabCap{#2}{#3}
  \begin{center} \TableFont \begin{tabular}{#1} #4 
  \end{tabular}\end{center}\end{table}}

\newenvironment{references}%
{
\footnotesize \frenchspacing
\renewcommand{\thesection}{}
\renewcommand{\in}{{\rm in }}
\renewcommand{\AA}{Astron.\ Astrophys.}
\newcommand{\AAS}{Astron.~Astrophys.~Suppl.~Ser.}
\newcommand{\ApJ}{Astrophys.\ J.}
\newcommand{\ApJS}{Astrophys.\ J.~Suppl.~Ser.}
\newcommand{\ApJL}{Astrophys.\ J.~Letters}
\newcommand{\AJ}{Astron.\ J.}
\newcommand{\IBVS}{IBVS}
\newcommand{\PASP}{P.A.S.P.}
\newcommand{\Acta}{Acta Astron.}
\newcommand{\MNRAS}{MNRAS}
\renewcommand{\and}{{\rm and }}
\section{{\rm REFERENCES}}
\sloppy \hyphenpenalty10000
\begin{list}{}{\leftmargin1cm\listparindent-1cm
\itemindent\listparindent\parsep0pt\itemsep0pt}}%
{\end{list}\vspace{2mm}}

\def\TYLDA{~}
\newlength{\DW}
\settowidth{\DW}{0}
\newcommand{\dw}{\hspace{\DW}}

\newcommand{\refitem}[5]{\item[]{#1} #2%
\def\REFARG{#3}\ifx\REFARG\TYLDA\else, {\it#3}\fi
\def\REFARG{#4}\ifx\REFARG\TYLDA\else, {\bf#4}\fi
\def\REFARG{#5}\ifx\REFARG\TYLDA\else, {#5}\fi.}

\newcommand{\Section}[1]{\section{#1}}
\newcommand{\Subsection}[1]{\subsection{#1}}
\newcommand{\Acknow}[1]{\par\vspace{5mm}{\bf Acknowledgements.} #1}
\pagestyle{myheadings}

\def\thefootnote{\fnsymbol{footnote}}
\begin{center}
{\large\bf The Optical Gravitational Lensing Experiment.\\
%\vskip3pt
Catalog of Microlensing Events in the Galactic Bulge\footnote{Based
on  observations obtained with the 1.3~m Warsaw telescope at the Las
Campanas  Observatory of the Carnegie Institution of Washington.}}
\vskip0.8cm
{\bf
A.~~U~d~a~l~s~k~i$^1$,~~K.~~\.Z~e~b~r~u~\'n$^1$
~~M.~~S~z~y~m~a~{\'n}~s~k~i$^1$,~~M.~~K~u~b~i~a~k$^1$,
~~G.~~P~i~e~t~r~z~y~\'n~s~k~i$^1$,~~I.~~S~o~s~z~y~{\'n}~s~k~i$^1$,
~~and ~~P.~~W~o~\'z~n~i~a~k$^2$}
\vskip3mm
{$^1$Warsaw University Observatory, Al.~Ujazdowskie~4, 00-478~Warszawa, Poland\\
e-mail: (udalski,zebrun,msz,mk,pietrzyn,soszynsk)@astrouw.edu.pl\\
$^2$ Princeton University Observatory, Princeton, NJ 08544-1001, USA\\
e-mail: wozniak@astro.princeton.edu}
\end{center}

\Abstract{

We present the Catalog of microlensing events detected toward the
Galactic bulge in three observing seasons, 1997--1999, during the
OGLE-II microlensing survey. The search for microlensing events was
performed using a database of about $4\cdot10^9$ photometric
measurements of about 20.5 million stars from the Galactic bulge. The
Catalog comprises 214 cases of microlensing events found in the fields
covering about 11 square degrees on the sky and distributed in different
parts of the Galactic bulge. The sample includes 20 binary microlensing
events, 14 of them are caustic crossing. In one case a double star is
likely lensed.

We present distribution of the basic parameters of microlensing events
and show preliminary rate of microlensing in different regions of the
Galactic bulge. The latter reveals clear dependence on the Galactic
coordinates. The dependence on $l$ indicates that the majority of lenses
toward the Galactic bulge are located in the Galactic bar. Models of the
Galactic bar seem to reasonably predict the observed spatial
distribution of microlensing events in the Galactic bulge.

All data presented in the Catalog and photometry of all events are
available from the OGLE Internet archive.
}{~}

\Section{Introduction}

During the past couple of years microlensing proved to be  a new and
potentially very powerful tool of modern astrophysics. Originally
proposed by Paczy{\'n}ski (1986, 1991) as a method of searching for dark
matter in the Galaxy it has been used for such different applications as
searching for planets, determination of parameters of stellar
atmospheres, studies of Galactic structure and many others.

After the original reports on discovery of the first cases of
microlensing in September 1993 (toward the LMC: MACHO survey -- Alcock
\etal 1993, EROS survey -- Aubourg \etal 1993; toward the Galactic
bulge:  OGLE survey -- Udalski \etal 1993) much observing work was done
to convince the astronomical community on the potentials of the newly
discovered class of events. Soon more cases of classical microlensing in
the Galactic bulge were announced (Udalski \etal 1994a, Alcock \etal
1995a), first cases of "exotic microlensing" like binary microlensing
(Udalski \etal 1994b) or events with parallax effect (Alcock \etal
1995b) were found. Also first estimates of the observed optical depth to
the Galactic bulge were published (Udalski \etal 1994c, Alcock \etal
1995a, 1997a) indicating that it is much larger than that predicted from
modeling. Many theoretical interpretations of these intriguing results
followed (see review by Paczy{\'n}ski 1996). First interpretation of
results of observations toward the LMC was also published (Alcock \etal
1996a).

Another important step was development and implementation of the so
called alert systems (the Early Warning System, EWS, of the OGLE survey
-- Udalski \etal 1994d and MACHO Alert system -- Alcock \etal 1996b)
which allow to detect the microlensing phenomena when an event is in
progress. This step changed the observing strategy of microlensing
surveys. The ability of detection in real time  made it possible to
perform follow-up observations, both photometric and spectroscopic, of
many events. New kind of microlensing studies, "follow-up" projects
concentrated on high time resolution observations of events discovered
by survey projects, were formed (\eg PLANET -- Albrow \etal 1998; MPS --
Rhie \etal 1999). From 1995 on, the vast majority of microlensing events
have been detected by the alert systems.\footnote{\noindent Information
on microlensing events in progress can be found:\newline OGLE-EWS: {\it
http://www.astrouw.edu.pl/\~{}ogle} \newline MACHO (1995--1999): {\it
http://darkstar.astro.washington.edu/} \newline EROS: {\it
http://www-dapnia.cea.fr/Spp/Experiences/EROS/alertes.html} \newline
PLANET: {\it http://www.astro.rug.nl/\~{}planet/} \newline MPS: {\it
http://bustard.phys.nd.edu/MPS/}
}

The microlensing field of astrophysics matured rapidly and entered the
second phase of extensive observations to largely increase statistic of
collected microlensing events. Both EROS and OGLE surveys considerably
increased their observing capabilities in 1996. First microlensing
events were discovered in other lines of sight -- toward the SMC (MACHO
-- Alcock \etal 1997b) and in the Galactic disk (EROS -- Derue \etal
1999) even as far as $70^{\circ}$ from the Galactic center (OGLE -- Mao
1999). Many interesting results were published by follow-up teams
(PLANET -- Albrow \etal 2000a,b, MPS -- Bennett \etal 1999a). Up to now
one can estimate the total number of registered microlensing events to
be about 500.

In this paper we present the catalog of microlensing event candidates
detected during the second phase of the Optical Gravitational Lensing
Experiment (OGLE-II) in three observing seasons 1997--1999. Although
during the OGLE-II phase observations are conducted in many lines of
sight only two events were discovered in the directions other than the
Galactic bulge so far.\footnote{1999-CAR-01 and 1999-LMC-01 see {\it
http://www.astrouw.edu.pl/\~{}ogle} for more information on these
events.} Therefore we decided to limit our catalog to the Galactic bulge
events only. Presented microlensing events were extracted from the
OGLE-II photometric databases with a technique similar to that applied
in the EWS alert system. Great care was paid to achieve the highest
possible completeness of the catalog. The Catalog comprises 214 cases of
microlensing toward the Galactic bulge.

The main goal of the paper is to provide the astronomical microlensing
community with the large set of photometric data of microlensing events
for further analysis. Photometry of all objects is available from the
OGLE Internet archive (see Section~5). Beside of the typical single mass
microlensing cases the Catalog includes several cases of "exotic
microlensing", like binary microlensing events etc.

The sample of presented events is already large enough that in spite of
possible incompleteness the distribution  of microlensing parameters can
be studied. In particular we present the first preliminary distribution
of the rate of microlensing in different parts of the Galactic bulge.
Studies of such a distribution can provide important constraints on the
origin of the Galactic bulge microlensing (bar \vs disk) and in general
on the structure of the Galaxy.

\Section{Observational Data}

All observations presented in this paper were carried out during the
second  phase of the OGLE experiment with the 1.3-m Warsaw telescope at
the Las  Campanas Observatory, Chile, which is operated by the Carnegie
Institution of  Washington. The telescope was equipped with a "first
generation" camera with a SITe ${2048\times2048}$ back-illuminated CCD
detector working in drift-scan mode. The pixel size was 24~$\mu$m giving
the scale of 0.417 arcsec/pixel. Observations of  the Galactic bulge 
were performed in the "medium" speed reading mode of CCD detector with
the  gain 7.1~e$^-$/ADU and readout noise of about 6.3~e$^-$. Details of
the  instrumentation setup can be found in Udalski, Kubiak and
Szyma{\'n}ski  (1997). 

\renewcommand{\TableFont}{\scriptsize}

\MakeTableSep{lccrrr}{12.5cm}{OGLE-II fields in the Galactic  bulge}
{
\hline
\noalign{\vskip3pt}
\multicolumn{1}{c}{Field} & RA (J2000)  & DEC (J2000) & \multicolumn{1}{c}{$l$}
& \multicolumn{1}{c}{$b$} & \multicolumn{1}{c}{$N_{\rm obs}$}\\
\hline
\noalign{\vskip3pt}
 BUL$\_$SC1  & 18\uph02\upm32\zdot\ups5 & $-29\arcd57\arcm41\arcs$ &   $ 1\zdot\arcd08$ & $-3\zdot\arcd62$ & 188\\
 BUL$\_$SC2  & 18\uph04\upm28\zdot\ups6 & $-28\arcd52\arcm35\arcs$ &   $ 2\zdot\arcd23$ & $-3\zdot\arcd46$ & 183\\
 BUL$\_$SC3  & 17\uph53\upm34\zdot\ups4 & $-29\arcd57\arcm56\arcs$ &   $ 0\zdot\arcd11$ & $-1\zdot\arcd93$ & 291\\
 BUL$\_$SC4  & 17\uph54\upm35\zdot\ups7 & $-29\arcd43\arcm41\arcs$ &   $ 0\zdot\arcd43$ & $-2\zdot\arcd01$ & 301\\
 BUL$\_$SC5  & 17\uph50\upm21\zdot\ups7 & $-29\arcd56\arcm49\arcs$ &   $-0\zdot\arcd23$ & $-1\zdot\arcd33$ & 288\\
 BUL$\_$SC6  & 18\uph08\upm03\zdot\ups7 & $-32\arcd07\arcm48\arcs$ &   $-0\zdot\arcd25$ & $-5\zdot\arcd70$ & 212\\
 BUL$\_$SC7  & 18\uph09\upm10\zdot\ups6 & $-32\arcd07\arcm40\arcs$ &   $-0\zdot\arcd14$ & $-5\zdot\arcd91$ & 207\\
 BUL$\_$SC8  & 18\uph23\upm06\zdot\ups2 & $-21\arcd47\arcm53\arcs$ &   $10\zdot\arcd48$ & $-3\zdot\arcd78$ & 198\\
 BUL$\_$SC9  & 18\uph24\upm02\zdot\ups5 & $-21\arcd47\arcm55\arcs$ &   $10\zdot\arcd59$ & $-3\zdot\arcd98$ & 201\\
 BUL$\_$SC10 & 18\uph20\upm06\zdot\ups6 & $-22\arcd23\arcm03\arcs$ &   $ 9\zdot\arcd64$ & $-3\zdot\arcd44$ & 207\\
 BUL$\_$SC11 & 18\uph21\upm06\zdot\ups5 & $-22\arcd23\arcm05\arcs$ &   $ 9\zdot\arcd74$ & $-3\zdot\arcd64$ & 205\\
 BUL$\_$SC12 & 18\uph16\upm06\zdot\ups3 & $-23\arcd57\arcm54\arcs$ &   $ 7\zdot\arcd80$ & $-3\zdot\arcd37$ & 192\\
 BUL$\_$SC13 & 18\uph17\upm02\zdot\ups6 & $-23\arcd57\arcm44\arcs$ &   $ 7\zdot\arcd91$ & $-3\zdot\arcd58$ & 193\\
 BUL$\_$SC14 & 17\uph47\upm02\zdot\ups7 & $-23\arcd07\arcm30\arcs$ &   $ 5\zdot\arcd23$ & $ 2\zdot\arcd81$ & 192\\
 BUL$\_$SC15 & 17\uph48\upm06\zdot\ups9 & $-23\arcd06\arcm09\arcs$ &   $ 5\zdot\arcd38$ & $ 2\zdot\arcd63$ & 189\\
 BUL$\_$SC16 & 18\uph10\upm06\zdot\ups7 & $-26\arcd18\arcm05\arcs$ &   $ 5\zdot\arcd10$ & $-3\zdot\arcd29$ & 185\\
 BUL$\_$SC17 & 18\uph11\upm03\zdot\ups6 & $-26\arcd12\arcm35\arcs$ &   $ 5\zdot\arcd28$ & $-3\zdot\arcd45$ & 179\\
 BUL$\_$SC18 & 18\uph07\upm03\zdot\ups5 & $-27\arcd12\arcm48\arcs$ &   $ 3\zdot\arcd97$ & $-3\zdot\arcd14$ & 195\\
 BUL$\_$SC19 & 18\uph08\upm02\zdot\ups4 & $-27\arcd12\arcm45\arcs$ &   $ 4\zdot\arcd08$ & $-3\zdot\arcd35$ & 186\\
 BUL$\_$SC20 & 17\uph59\upm19\zdot\ups1 & $-28\arcd52\arcm55\arcs$ &   $ 1\zdot\arcd68$ & $-2\zdot\arcd47$ & 228\\
 BUL$\_$SC21 & 18\uph00\upm22\zdot\ups3 & $-28\arcd51\arcm45\arcs$ &   $ 1\zdot\arcd80$ & $-2\zdot\arcd66$ & 221\\
 BUL$\_$SC22 & 17\uph56\upm47\zdot\ups6 & $-30\arcd47\arcm46\arcs$ &   $-0\zdot\arcd26$ & $-2\zdot\arcd95$ & 264\\
 BUL$\_$SC23 & 17\uph57\upm54\zdot\ups5 & $-31\arcd12\arcm36\arcs$ &   $-0\zdot\arcd50$ & $-3\zdot\arcd36$ & 246\\
 BUL$\_$SC24 & 17\uph53\upm17\zdot\ups9 & $-32\arcd52\arcm45\arcs$ &   $-2\zdot\arcd44$ & $-3\zdot\arcd36$ & 239\\
 BUL$\_$SC25 & 17\uph54\upm26\zdot\ups1 & $-32\arcd52\arcm49\arcs$ &   $-2\zdot\arcd32$ & $-3\zdot\arcd56$ & 238\\
 BUL$\_$SC26 & 17\uph47\upm15\zdot\ups5 & $-34\arcd59\arcm31\arcs$ &   $-4\zdot\arcd90$ & $-3\zdot\arcd37$ & 218\\
 BUL$\_$SC27 & 17\uph48\upm23\zdot\ups6 & $-35\arcd09\arcm32\arcs$ &   $-4\zdot\arcd92$ & $-3\zdot\arcd65$ & 209\\
 BUL$\_$SC28 & 17\uph47\upm05\zdot\ups8 & $-37\arcd07\arcm47\arcs$ &   $-6\zdot\arcd76$ & $-4\zdot\arcd42$ & 214\\
 BUL$\_$SC29 & 17\uph48\upm10\zdot\ups8 & $-37\arcd07\arcm21\arcs$ &   $-6\zdot\arcd64$ & $-4\zdot\arcd62$ & 206\\
 BUL$\_$SC30 & 18\uph01\upm25\zdot\ups0 & $-28\arcd49\arcm55\arcs$ &   $ 1\zdot\arcd94$ & $-2\zdot\arcd84$ & 228\\
 BUL$\_$SC31 & 18\uph02\upm22\zdot\ups6 & $-28\arcd37\arcm21\arcs$ &   $ 2\zdot\arcd23$ & $-2\zdot\arcd94$ & 231\\
 BUL$\_$SC32 & 18\uph03\upm26\zdot\ups8 & $-28\arcd38\arcm02\arcs$ &   $ 2\zdot\arcd34$ & $-3\zdot\arcd14$ & 223\\
 BUL$\_$SC33 & 18\uph05\upm30\zdot\ups9 & $-28\arcd52\arcm50\arcs$ &   $ 2\zdot\arcd35$ & $-3\zdot\arcd66$ & 179\\
 BUL$\_$SC34 & 17\uph58\upm18\zdot\ups5 & $-29\arcd07\arcm50\arcs$ &   $ 1\zdot\arcd35$ & $-2\zdot\arcd40$ & 218\\
 BUL$\_$SC35 & 18\uph04\upm28\zdot\ups6 & $-27\arcd56\arcm56\arcs$ &   $ 3\zdot\arcd05$ & $-3\zdot\arcd00$ & 169\\
 BUL$\_$SC36 & 18\uph05\upm31\zdot\ups2 & $-27\arcd56\arcm44\arcs$ &   $ 3\zdot\arcd16$ & $-3\zdot\arcd20$ & 199\\
 BUL$\_$SC37 & 17\uph52\upm32\zdot\ups2 & $-29\arcd57\arcm44\arcs$ &   $ 0\zdot\arcd00$ & $-1\zdot\arcd74$ & 299\\
 BUL$\_$SC38 & 18\uph01\upm28\zdot\ups0 & $-29\arcd57\arcm01\arcs$ &   $ 0\zdot\arcd97$ & $-3\zdot\arcd42$ & 186\\
 BUL$\_$SC39 & 17\uph55\upm39\zdot\ups1 & $-29\arcd44\arcm52\arcs$ &   $ 0\zdot\arcd53$ & $-2\zdot\arcd21$ & 310\\
 BUL$\_$SC40 & 17\uph51\upm06\zdot\ups1 & $-33\arcd15\arcm11\arcs$ &   $-2\zdot\arcd99$ & $-3\zdot\arcd14$ & 200\\
 BUL$\_$SC41 & 17\uph52\upm07\zdot\ups2 & $-33\arcd07\arcm41\arcs$ &   $-2\zdot\arcd78$ & $-3\zdot\arcd27$ & 201\\
 BUL$\_$SC42 & 18\uph09\upm05\zdot\ups0 & $-26\arcd51\arcm53\arcs$ &   $ 4\zdot\arcd48$ & $-3\zdot\arcd38$ & 195\\
 BUL$\_$SC43 & 17\uph35\upm13\zdot\ups5 & $-27\arcd11\arcm00\arcs$ &   $ 0\zdot\arcd37$ & $ 2\zdot\arcd95$ & 243\\
 BUL$\_$SC44 & 17\uph49\upm22\zdot\ups4 & $-30\arcd02\arcm45\arcs$ &   $-0\zdot\arcd43$ & $-1\zdot\arcd19$ & 251\\
 BUL$\_$SC45 & 18\uph03\upm36\zdot\ups5 & $-30\arcd05\arcm00\arcs$ &   $ 0\zdot\arcd98$ & $-3\zdot\arcd94$ &  88\\    
 BUL$\_$SC46 & 18\uph04\upm39\zdot\ups7 & $-30\arcd05\arcm11\arcs$ &   $ 1\zdot\arcd09$ & $-4\zdot\arcd14$ &  84\\
 BUL$\_$SC47 & 17\uph27\upm03\zdot\ups7 & $-39\arcd47\arcm16\arcs$ &  $-11\zdot\arcd19$ & $-2\zdot\arcd60$ & 147\\
 BUL$\_$SC48 & 17\uph28\upm14\zdot\ups0 & $-39\arcd46\arcm58\arcs$ &  $-11\zdot\arcd07$ & $-2\zdot\arcd78$ & 142\\
 BUL$\_$SC49 & 17\uph29\upm25\zdot\ups1 & $-40\arcd16\arcm21\arcs$ &  $-11\zdot\arcd36$ & $-3\zdot\arcd25$ & 142\\
\hline}

Regular observations of the Galactic bulge started on March~23, 1997.
The observing season of the Galactic bulge lasts typically from mid
February up to the end of October. In this paper we present microlensing
events detected during the seasons 1997--1999.

\begin{figure}[htb]
\psfig{figure=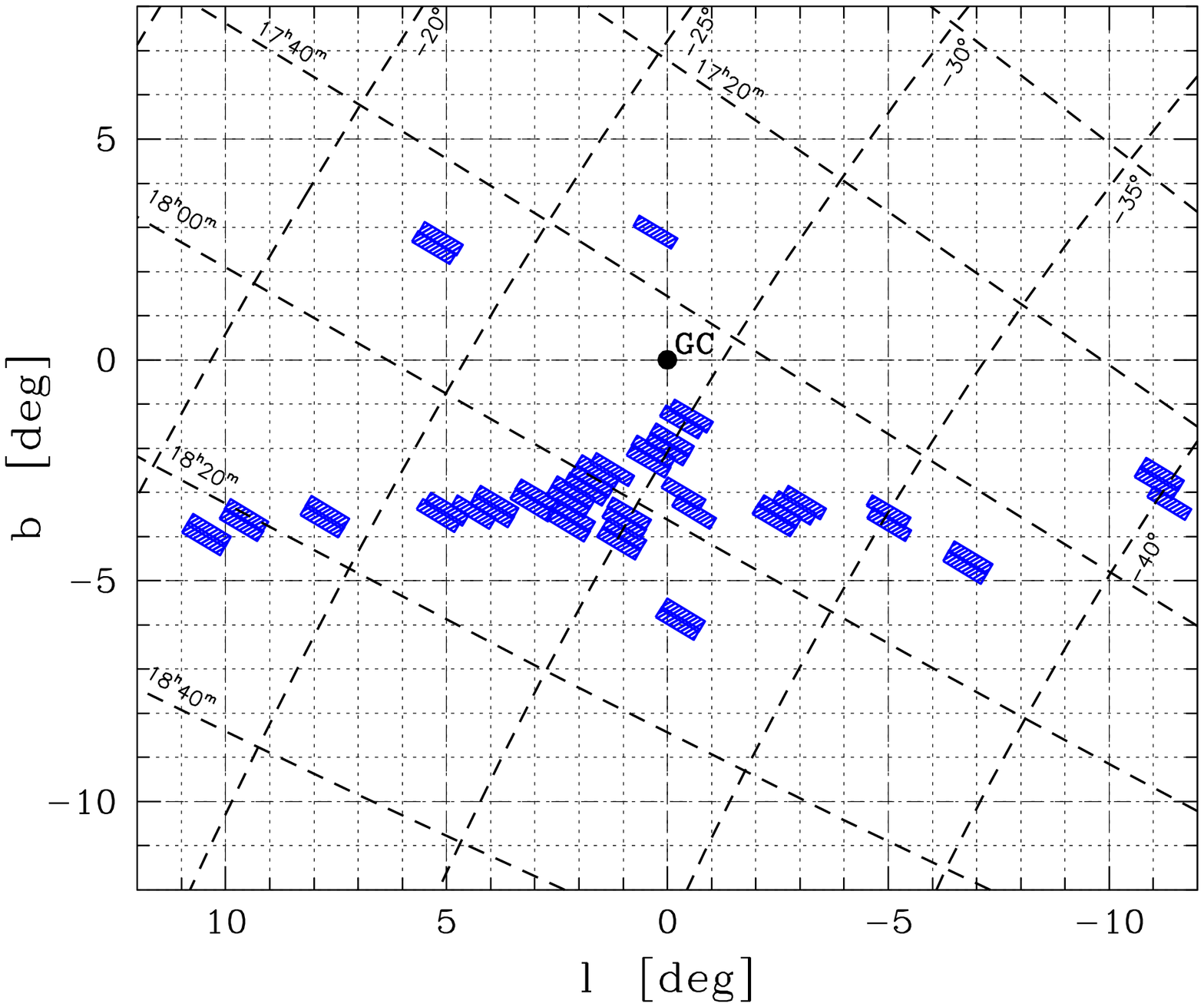,bbllx=15pt,bblly=150pt,bburx=565pt,bbury=615pt,width=12.5cm,clip=}
\FigCap{OGLE-II fields in the Galactic bulge.}
\end{figure}

Forty nine driftscan fields were observed, each covering $14.2\times
57$~arcmins on the sky, giving total area of about 11~square degrees.
Forty seven of these fields were monitored frequently for detection of
microlensing events. The two remaining fields (BUL$\_$SC45 and
BUL$\_$SC46) were observed only from time to time, mostly for
maintaining phasing of variable stars discovered during the OGLE-I phase
of the OGLE survey. Fields BUL$\_$SC47, BUL$\_$SC48 and BUL$\_$SC49 were
added to the list of targets in the second observing season, so 
observations of these fields span seasons  1998 and 1999 only.

Table~1 lists the equatorial and Galactic coordinates of the center of
each field and its acronym. Also the number of collected frames is
provided. Fig.~1 shows schematically location  of the Galactic fields in
($l$, $b$) and (RA, DEC) coordinate systems. Most fields are located
along the regions of smaller interstellar extinction, \ie large stellar
density at $b\approx-3\zdot\arcd5$ and spanning large range of the
Galactic longitude -- from $l=-11\arcd$ to $l=+11\arcd$. A few fields
are located on the other side of the Galactic equator at positive $b$.
In general the interstellar extinction there is much larger making
selection of dense stellar fields suitable for microlensing search  much
more difficult.

Observations of the Galactic bulge fields were obtained in the standard
{\it  BVI}-bands. The effective exposure time was 87, 124 and 162
seconds for the {\it I, V} and {\it B}-band, respectively.  The vast
majority of observations were done through the {\it I}-band filter while
images on only several epochs were collected in the {\it BV}-bands. The
instrumental system closely resembles the standard {\it BVI}. All
photometric data of the Galactic bulge will be ultimately calibrated and
released in the form of photometric maps (similar to the SMC data:
Udalski \etal 1998). In this paper we present only {\it I}-band
observations. The data are only preliminarily calibrated but uncertainty
of the zero points of photometry should not be larger than
$\pm0.05$~mag. The median seeing of the entire set of several thousand
collected frames of the Galactic bulge is equal to 1\zdot\arcs29.

\Section{Search Algorithm}

The Galactic bulge photometric data collected during the OGLE-II survey
are naturally divided into separate seasons. We used the {\it I}-band
photometric data from each season to create databases of non-variable
stars in every field. Then we compared brightness of selected
non-variable stars with that in other seasons, looking for objects which
vary in brightness.  Databases of non-variable stars were created for
each season, \ie 1997, 1998 and 1999 separately, and the search for
variable objects was performed independently with all possible
combinations of "non-variable" and "variable" seasons. In this manner we
were able to detect inter-seasonal events and verify most candidates by
comparison of results in a given season obtained with databases of
non-variable stars from two other seasons.

The star entered the database of non-variable objects in a given season
when the standard deviation of all observations collected during that
season was smaller than the maximum standard deviation of non-variable
stars at given magnitude. The latter was determined by analysis of the
distribution of the brightness standard deviation of all observed stars
in the field. We limited the databases to  objects brighter than
$I=19.5$~mag. The total number of non-variable stars searched for
microlensing phenomena in our fields was more than 20.5 millions (about
$4\cdot10^9$ photometric measurements).

Comparison of number of stars which were included into databases of
non-variable stars in different seasons provided  information  on the
actual number of real stellar objects. They were similar to within 5\%
indicating that only a small fraction of stars in these databases are
spurious objects like artifacts in saturated regions of bright stars
etc.

Candidates for variable objects were selected by comparison of their 
photometry with the mean magnitude in the template "non-variable"
season. When five consecutive observations were brighter than $3\sigma$
limit  over the "non-variable" magnitude, such an object was marked for
further analysis. The typical $3\sigma$ threshold was equal to 0.06~mag
for a star of $I<16$~mag, 0.20~mag for $I=18$~mag and 0.60~mag for the
limiting $I=19.5$~mag. At this stage we decided to visually inspect
light curves of all such candidates to avoid removing non-standard
events like for instance caustic crossing binary events. All candidates
which displayed more than one episode of brightening (non consistent
with microlensing type), turned out to be evident long-term variable
stars  and those which evidently exhibited spurious variability were
removed from the list of candidates. In the next step images taken at
the epochs of normal (constant) brightness and near the maximum of
brightening were inspected to verify whether the brightening is real. In
many cases the brightening was spurious due to proximity of large
amplitude variable star etc. Such objects were also removed.  We did not
checked achromaticity of the brightening. Because of sparse coverage in
the bands other than {\it I} it would be practically impossible to
perform that test for most of our candidates. On the other hand the
experience of the OGLE-I phase as well as experience of other teams
indicate that in the case of stellar population in the Galactic bulge
(contrary to the Magellanic Clouds) there are practically no objects
which would mimic microlensing event light curve. Thus, the
achromaticity is not that important to extract the real microlensing
events from the background of variable stars. Moreover, in the very
dense stellar fields of the Galactic bulge blending can cause quite
severe deviations from achromaticity making the latter condition not
very strong.

We were left with a list of 214 candidates for microlensing events which
passed all our tests. They are shown and analyzed in the next Sections.

\renewcommand{\arraystretch}{1.3}
\renewcommand{\TableFont}{\tiny}
\setcounter{table}{1}
\MakeTableSep{
c@{\hspace{4pt}}
r@{\hspace{4pt}}
c@{\hspace{4pt}}
c@{\hspace{4pt}}
c@{\hspace{4pt}}
r@{\hspace{3pt}}
r@{\hspace{3pt}}
r@{\hspace{4pt}}
r@{\hspace{4pt}}
l}{12.5cm}{Parameters of microlensing events}
{\hline
\noalign{\vskip3pt}
\multicolumn{2}{c}{Identification} & EWS & RA(2000) & DEC(2000) & 
\multicolumn{1}{c}{$T_{\rm max}$} & 
\multicolumn{1}{c}{$t_0$} & 
\multicolumn{1}{c}{$A_{\rm max}$} & \multicolumn{1}{c}{$I_0$} & Rem.\\
Field & \multicolumn{1}{c}{No} &&&&
\multicolumn{1}{c}{[HJD--2450000]} & \multicolumn{1}{c}{[days]} & &
\multicolumn{1}{c}{[mag]} &\\
\noalign{\vskip3pt}
\hline
\noalign{\vskip3pt}
BUL$\_$SC1  & 460673  &      --     & 18\uph02\upm44\zdot\ups86 & $-29\arcd58\arcm18\zdot\arcs2$ & $ 989.438\pm 0.849$ & $ 19.21\pm1.63$ & $ 2.24\pm 0.06$ & $19.063\pm0.010$ &\\  
BUL$\_$SC2  & 8398    &      --     & 18\uph04\upm05\zdot\ups69 & $-29\arcd20\arcm10\zdot\arcs0$ & $1064.358\pm 0.699$ & $ 19.65\pm2.10$ & $ 3.28\pm 0.26$ & $18.507\pm0.015$ &\\ 
BUL$\_$SC2  & 27414   &      --     & 18\uph04\upm04\zdot\ups92 & $-29\arcd11\arcm38\zdot\arcs4$ & $ 562.089\pm 0.130$ & $ 10.63\pm0.25$ & $ 3.01\pm 0.05$ & $18.253\pm0.005$ &\\ 
BUL$\_$SC2  & 65831   & 1999-BUL-44 & 18\uph04\upm06\zdot\ups85 & $-29\arcd01\arcm17\zdot\arcs2$ & $1460.255\pm 0.633$ & $ 32.31\pm0.60$ & $ 1.35\pm 0.00$ & $14.604\pm0.001$ &\\ 
BUL$\_$SC2  & 495787  & 1998-BUL-37 & 18\uph04\upm42\zdot\ups38 & $-28\arcd59\arcm38\zdot\arcs7$ & $1082.555\pm 0.481$ & $ 18.96\pm1.34$ & $ 2.49\pm 0.10$ & $18.616\pm0.007$ &\\ 
BUL$\_$SC3  & 77887   & 1999-BUL-27 & 17\uph53\upm09\zdot\ups83 & $-30\arcd02\arcm52\zdot\arcs5$ & $1366.209\pm 0.425$ & $ 25.80\pm0.57$ & $ 1.58\pm 0.01$ & $17.220\pm0.002$ &\\ 
BUL$\_$SC3  & 91382   & 1998-BUL-14 & 17\uph53\upm09\zdot\ups33 & $-30\arcd01\arcm12\zdot\arcs1$ & $ 956.003\pm 0.046$ & $ 38.31\pm0.24$ & $14.55\pm 0.29$ & $16.544\pm0.002$ &\\ 
BUL$\_$SC3  & 457512  & 1999-BUL-14 & 17\uph53\upm41\zdot\ups47 & $-30\arcd14\arcm54\zdot\arcs0$ & $1321.087\pm 1.040$ & $ 20.56\pm1.40$ & $ 1.36\pm 0.02$ & $17.385\pm0.004$ &\\ 
BUL$\_$SC3  & 469296  & 1998-BUL-01 & 17\uph53\upm40\zdot\ups36 & $-30\arcd10\arcm20\zdot\arcs1$ & $ 887.186\pm 0.455$ & $ 39.67\pm1.03$ & $ 3.26\pm 0.06$ & $17.222\pm0.003$ &\\ 
BUL$\_$SC3  & 480386  &      --     & 17\uph53\upm41\zdot\ups54 & $-30\arcd08\arcm12\zdot\arcs8$ & $ 611.535\pm 0.131$ & $  9.65\pm0.44$ & $ 2.04\pm 0.05$ & $17.897\pm0.004$ &\\ 
BUL$\_$SC3  & 541151  &      --     & 17\uph53\upm47\zdot\ups52 & $-29\arcd47\arcm48\zdot\arcs7$ & $ 690.255\pm 0.268$ & $ 29.50\pm0.81$ & $ 3.50\pm 0.07$ & $18.041\pm0.006$ & BIN?\\         
BUL$\_$SC3  & 576464  &      --     & 17\uph53\upm41\zdot\ups56 & $-29\arcd39\arcm06\zdot\arcs0$ & $ 656.642\pm 0.329$ & $ 31.75\pm0.51$ & $ 1.74\pm 0.01$ & $16.271\pm0.002$ &\\ 
BUL$\_$SC3  & 577547  &      --     & 17\uph53\upm37\zdot\ups83 & $-29\arcd38\arcm22\zdot\arcs7$ & $ 950.800\pm 0.553$ & $  3.06\pm0.92$ & $ 1.37\pm 0.05$ & $17.109\pm0.003$ &\\ 
BUL$\_$SC3  & 588309  & 1998-BUL-27 & 17\uph53\upm38\zdot\ups65 & $-29\arcd33\arcm42\zdot\arcs3$ & $1048.480\pm 0.686$ & $ 23.22\pm0.78$ & $ 1.20\pm 0.00$ & $14.906\pm0.001$ &\\ 
BUL$\_$SC3  & 590098  &      --     & 17\uph53\upm47\zdot\ups66 & $-29\arcd34\arcm45\zdot\arcs6$ &           --        &         --      &       --        & $17.627\pm0.006$ & BCC\\
BUL$\_$SC3  & 601945  & 1998-BUL-08 & 17\uph53\upm36\zdot\ups63 & $-29\arcd31\arcm21\zdot\arcs2$ & $ 924.120\pm 0.811$ & $ 42.55\pm1.31$ & $ 1.61\pm 0.02$ & $17.405\pm0.002$ &\\ 
BUL$\_$SC3  & 792295  & 1998-BUL-28 & 17\uph53\upm54\zdot\ups15 & $-29\arcd36\arcm40\zdot\arcs6$ &           --        &         --      &       --        & $19.283\pm0.026$ & BCC\\
BUL$\_$SC4  & 134300  & 1998-BUL-26 & 17\uph54\upm13\zdot\ups04 & $-29\arcd35\arcm14\zdot\arcs1$ & $1039.554\pm 0.692$ & $ 22.41\pm0.92$ & $ 1.42\pm 0.01$ & $16.784\pm0.002$ &\\ 
BUL$\_$SC4  & 201173  & 1999-BUL-21 & 17\uph54\upm15\zdot\ups99 & $-29\arcd18\arcm49\zdot\arcs7$ & $1318.917\pm 0.261$ & $ 10.50\pm0.88$ & $ 2.92\pm 0.09$ & $18.953\pm0.012$ &\\ 
BUL$\_$SC4  & 267762  & 1998-BUL-18 & 17\uph54\upm21\zdot\ups79 & $-29\arcd53\arcm24\zdot\arcs0$ & $ 971.029\pm 0.051$ & $  6.96\pm0.08$ & $ 4.10\pm 0.03$ & $15.507\pm0.001$ &\\ 
BUL$\_$SC4  & 273327  &      --     & 17\uph54\upm19\zdot\ups53 & $-29\arcd51\arcm49\zdot\arcs7$ & $ 648.111\pm 0.118$ & $  7.88\pm0.38$ & $ 4.31\pm 0.36$ & $18.405\pm0.007$ &\\ 
BUL$\_$SC4  & 411329  &      --     & 17\uph54\upm43\zdot\ups67 & $-30\arcd08\arcm31\zdot\arcs4$ & $ 667.577\pm 0.121$ & $  3.68\pm0.23$ & $ 2.54\pm 0.14$ & $18.208\pm0.005$ &\\ 
BUL$\_$SC4  & 451457  & 1998-BUL-39 & 17\uph54\upm44\zdot\ups69 & $-29\arcd55\arcm42\zdot\arcs8$ & $1129.430\pm 0.768$ & $ 37.91\pm1.13$ & $ >5.3        $ & $16.899\pm0.002$ &\\ 
BUL$\_$SC4  & 463924  & 1998-BUL-31 & 17\uph54\upm43\zdot\ups94 & $-29\arcd51\arcm18\zdot\arcs2$ & $1061.039\pm 0.681$ & $ 30.20\pm1.10$ & $ 1.44\pm 0.01$ & $16.623\pm0.002$ &\\ 
BUL$\_$SC4  & 478016  &      --     & 17\uph54\upm48\zdot\ups44 & $-29\arcd48\arcm39\zdot\arcs1$ & $1279.820\pm 0.658$ & $ 21.64\pm0.88$ & $ 1.73\pm 0.04$ & $17.588\pm0.003$ &\\ 
BUL$\_$SC4  & 522952  & 1999-BUL-02 & 17\uph54\upm38\zdot\ups64 & $-29\arcd33\arcm12\zdot\arcs8$ & $1257.217\pm 0.176$ & $ 23.09\pm0.36$ & $ 2.38\pm 0.03$ & $15.440\pm0.001$ &\\ 
BUL$\_$SC4  & 568740  & 1998-BUL-06 & 17\uph54\upm49\zdot\ups35 & $-29\arcd20\arcm25\zdot\arcs0$ & $ 915.902\pm 0.436$ & $ 16.40\pm0.43$ & $ 1.43\pm 0.02$ & $15.637\pm0.001$ &\\ 
BUL$\_$SC4  & 624085  &      --     & 17\uph55\upm00\zdot\ups19 & $-29\arcd59\arcm53\zdot\arcs6$ & $ 659.490\pm 0.471$ & $ 19.23\pm0.48$ & $ 1.19\pm 0.00$ & $15.474\pm0.001$ &\\ 
BUL$\_$SC4  & 631094  &      --     & 17\uph54\upm58\zdot\ups82 & $-30\arcd00\arcm11\zdot\arcs7$ & $1037.156\pm 0.507$ & $ 18.40\pm1.06$ & $ 2.58\pm 0.07$ & $18.938\pm0.007$ &\\ 
BUL$\_$SC4  & 681764  & 1999-BUL-29 & 17\uph54\upm55\zdot\ups29 & $-29\arcd44\arcm38\zdot\arcs6$ & $1364.788\pm 0.295$ & $ 27.36\pm1.06$ & $ 3.99\pm 0.11$ & $18.884\pm0.012$ &\\ 
BUL$\_$SC4  & 708424  & 1999-BUL-41 & 17\uph55\upm00\zdot\ups07 & $-29\arcd35\arcm03\zdot\arcs7$ & $1397.787\pm 0.026$ & $  5.96\pm0.06$ & $ 4.09\pm 0.04$ & $15.670\pm0.001$ &\\ 
BUL$\_$SC5  & 59256   & 1998-BUL-30 & 17\uph49\upm55\zdot\ups59 & $-29\arcd48\arcm03\zdot\arcs3$ & $1043.465\pm 0.112$ & $ 12.69\pm0.44$ & $ 3.68\pm 0.07$ & $18.845\pm0.006$ &\\ 
BUL$\_$SC5  & 63662   & 1998-BUL-16 & 17\uph49\upm52\zdot\ups76 & $-29\arcd45\arcm27\zdot\arcs2$ & $ 934.260\pm 0.384$ & $ 15.72\pm0.93$ & $ 2.33\pm 0.05$ & $17.865\pm0.003$ &\\ 
BUL$\_$SC5  & 152125  &      --     & 17\uph50\upm13\zdot\ups30 & $-29\arcd49\arcm57\zdot\arcs6$ &          --         &        --       &       --        & $15.123\pm0.002$ &\\ 
BUL$\_$SC5  & 244353  &      --     & 17\uph50\upm36\zdot\ups09 & $-30\arcd01\arcm46\zdot\arcs6$ &          --         &        --       &       --        &        --        & VAR?\\
BUL$\_$SC5  & 372272  &      --     & 17\uph50\upm42\zdot\ups73 & $-29\arcd58\arcm13\zdot\arcs8$ & $ 617.674\pm 0.343$ & $ 16.03\pm0.48$ & $ 1.18\pm 0.00$ & $15.583\pm0.001$ &\\ 
BUL$\_$SC6  & 243857  & 1999-BUL-03 & 18\uph08\upm02\zdot\ups64 & $-31\arcd49\arcm05\zdot\arcs1$ & $1252.014\pm 0.093$ & $ 85.09\pm0.81$ & $12.72\pm 0.14$ & $17.909\pm0.006$ &\\ 
BUL$\_$SC7  & 350881  & 1999-BUL-38 & 18\uph09\upm34\zdot\ups21 & $-32\arcd33\arcm00\zdot\arcs1$ & $1407.060\pm 0.926$ & $ 40.99\pm1.26$ & $ 1.63\pm 0.01$ & $17.725\pm0.003$ &\\ 
BUL$\_$SC10 & 220492  &      --     & 18\uph20\upm06\zdot\ups49 & $-22\arcd03\arcm58\zdot\arcs6$ & $ 584.971\pm 0.808$ & $108.26\pm2.08$ & $ 1.84\pm 0.01$ & $17.578\pm0.003$ &\\ 
BUL$\_$SC10 & 294229  &      --     & 18\uph20\upm20\zdot\ups60 & $-22\arcd24\arcm10\zdot\arcs1$ & $ 625.704\pm 0.166$ & $ 33.09\pm0.75$ & $ 4.86\pm 0.07$ & $18.950\pm0.010$ &\\ 
BUL$\_$SC10 & 454300  & 1999-BUL-36 & 18\uph20\upm30\zdot\ups00 & $-21\arcd57\arcm04\zdot\arcs7$ & $1392.580\pm 0.071$ & $ 28.30\pm0.22$ & $15.84\pm 0.78$ & $17.687\pm0.003$ &\\ 
BUL$\_$SC13 & 148843  & 1999-BUL-39 & 18\uph16\upm53\zdot\ups06 & $-24\arcd25\arcm47\zdot\arcs3$ & $1437.173\pm 0.850$ & $ 49.28\pm1.92$ & $ 2.01\pm 0.03$ & $17.997\pm0.007$ &\\ 
BUL$\_$SC14 & 177997  & 1999-BUL-09 & 17\uph47\upm02\zdot\ups44 & $-23\arcd30\arcm34\zdot\arcs8$ & $1287.744\pm 0.111$ & $ 14.68\pm0.32$ & $ >12.4       $ & $18.930\pm0.007$ &\\ 
BUL$\_$SC14 & 185869  &      --     & 17\uph47\upm00\zdot\ups43 & $-23\arcd27\arcm54\zdot\arcs4$ & $ 688.055\pm 1.346$ & $ 18.28\pm1.53$ & $ 1.37\pm 0.03$ & $17.713\pm0.004$ &\\ 
BUL$\_$SC14 & 194722  & 1998-BUL-24 & 17\uph46\upm47\zdot\ups64 & $-23\arcd24\arcm02\zdot\arcs0$ & $ 995.374\pm 1.482$ & $ 13.94\pm1.20$ & $ 1.14\pm 0.01$ & $16.036\pm0.001$ &\\ 
BUL$\_$SC14 & 358001  & 1999-BUL-13 & 17\uph47\upm05\zdot\ups18 & $-23\arcd19\arcm46\zdot\arcs1$ & $1317.959\pm 0.130$ & $ 18.34\pm0.19$ & $ 1.73\pm 0.01$ & $15.564\pm0.001$ &\\ 
BUL$\_$SC14 & 463284  &      --     & 17\uph47\upm02\zdot\ups91 & $-22\arcd40\arcm26\zdot\arcs3$ & $ 591.477\pm 1.761$ & $ 45.39\pm4.00$ & $ 1.53\pm 0.03$ & $18.827\pm0.011$ &\\ 
BUL$\_$SC14 & 480480  &      --     & 17\uph47\upm26\zdot\ups02 & $-23\arcd30\arcm03\zdot\arcs6$ & $ 574.909\pm 0.244$ & $ 13.69\pm0.36$ & $ 1.42\pm 0.01$ & $16.617\pm0.002$ &\\ 
BUL$\_$SC14 & 508950  &      --     & 17\uph47\upm20\zdot\ups66 & $-23\arcd20\arcm08\zdot\arcs8$ & $ 802.674\pm 1.929$ & $ 90.20\pm4.21$ & $ 2.36\pm 0.45$ & $17.965\pm0.005$ &\\ 
BUL$\_$SC15 & 20039   &      --     & 17\uph47\upm37\zdot\ups22 & $-23\arcd27\arcm03\zdot\arcs0$ & $ 675.003\pm 0.132$ & $ 41.65\pm0.52$ & $27.76\pm 7.07$ & $17.540\pm0.006$ &\\ 
BUL$\_$SC15 & 165930  &      --     & 17\uph48\upm01\zdot\ups11 & $-23\arcd31\arcm36\zdot\arcs5$ & $ 664.956\pm 0.469$ & $ 11.77\pm0.74$ & $ 1.71\pm 0.03$ & $18.039\pm0.004$ &\\ 
BUL$\_$SC15 & 345180  &      --     & 17\uph48\upm13\zdot\ups79 & $-23\arcd20\arcm38\zdot\arcs0$ & $ 602.505\pm 0.886$ & $ 36.38\pm1.39$ & $ 1.18\pm 0.00$ & $15.521\pm0.002$ &\\ 
BUL$\_$SC15 & 373196  &      --     & 17\uph48\upm07\zdot\ups05 & $-23\arcd12\arcm03\zdot\arcs1$ &           --        &         --      &        --       & $19.339\pm0.032$ & BCC\\
\hline
}

\setcounter{table}{1}
\MakeTableSep{
c@{\hspace{4pt}}
r@{\hspace{4pt}}
c@{\hspace{4pt}}
c@{\hspace{4pt}}
c@{\hspace{4pt}}
r@{\hspace{3pt}}
r@{\hspace{3pt}}
r@{\hspace{4pt}}
r@{\hspace{4pt}}
l}{12.5cm}{Continued}
{\hline
\noalign{\vskip3pt}
\multicolumn{2}{c}{Identification} & EWS & RA(2000) & DEC(2000) & 
\multicolumn{1}{c}{$T_{\rm max}$} & 
\multicolumn{1}{c}{$t_0$} & 
\multicolumn{1}{c}{$A_{\rm max}$} & \multicolumn{1}{c}{$I_0$} & Rem.\\
Field & \multicolumn{1}{c}{No} &&&&
\multicolumn{1}{c}{[HJD--2450000]} & \multicolumn{1}{c}{[days]} & &
\multicolumn{1}{c}{[mag]} &\\
\noalign{\vskip3pt}
\hline
\noalign{\vskip3pt}
BUL$\_$SC15 & 418332  &      --     & 17\uph48\upm08\zdot\ups12 & $-22\arcd53\arcm12\zdot\arcs9$ & $1359.106\pm 1.588$ & $ 51.61\pm2.51$ & $ 1.46\pm 0.02$ & $18.079\pm0.006$ &\\ 
BUL$\_$SC15 & 530219  &      --     & 17\uph48\upm29\zdot\ups61 & $-23\arcd04\arcm54\zdot\arcs7$ & $ 934.847\pm 1.756$ & $ 49.19\pm1.92$ & $ 1.17\pm 0.00$ & $14.989\pm0.001$ &\\ 
BUL$\_$SC15 & 539556  &      --     & 17\uph48\upm27\zdot\ups68 & $-23\arcd00\arcm11\zdot\arcs8$ & $ 665.035\pm 0.384$ & $ 34.77\pm0.54$ & $ 1.67\pm 0.01$ & $16.130\pm0.002$ &\\ 
BUL$\_$SC16 & 32304   &      --     & 18\uph09\upm38\zdot\ups48 & $-26\arcd32\arcm26\zdot\arcs7$ &           --        &         --      &        --       & $18.540\pm0.020$ & BCC\\
BUL$\_$SC16 & 436041  &      --     & 18\uph10\upm10\zdot\ups17 & $-26\arcd21\arcm03\zdot\arcs2$ & $ 540.903\pm 1.712$ & $ 70.71\pm2.87$ & $ 5.31\pm 0.47$ & $18.482\pm0.007$ &\\ 
BUL$\_$SC17 & 270411  & 1998-BUL-33 & 18\uph10\upm59\zdot\ups47 & $-26\arcd13\arcm16\zdot\arcs3$ & $1043.208\pm 0.215$ & $ 18.53\pm0.61$ & $ 3.44\pm 0.09$ & $18.518\pm0.006$ &\\ 
BUL$\_$SC17 & 509249  & 1999-BUL-24 & 18\uph11\upm10\zdot\ups91 & $-25\arcd50\arcm00\zdot\arcs5$ & $1336.000\pm 0.581$ & $  9.71\pm0.90$ & $ 1.79\pm 0.06$ & $18.274\pm0.008$ &\\ 
BUL$\_$SC18 & 2397    &      --     & 18\uph06\upm34\zdot\ups59 & $-27\arcd39\arcm51\zdot\arcs4$ & $ 626.890\pm 0.096$ & $ 19.30\pm0.51$ & $ 5.11\pm 0.08$ & $18.051\pm0.008$ &\\ 
BUL$\_$SC18 & 81415   & 1999-BUL-30 & 18\uph06\upm37\zdot\ups16 & $-27\arcd16\arcm26\zdot\arcs9$ & $1358.750\pm 0.576$ & $ 12.89\pm1.08$ & $ 1.86\pm 0.08$ & $18.547\pm0.010$ &\\ 
BUL$\_$SC18 & 242825  &      --     & 18\uph06\upm54\zdot\ups84 & $-27\arcd29\arcm21\zdot\arcs7$ & $ 996.149\pm 1.751$ & $ 35.56\pm1.90$ & $ 1.50\pm 0.03$ & $17.740\pm0.004$ &\\ 
BUL$\_$SC18 & 424029  &      --     & 18\uph07\upm05\zdot\ups61 & $-27\arcd32\arcm39\zdot\arcs4$ & $1359.145\pm 0.459$ & $ 11.28\pm0.82$ & $ 1.49\pm 0.04$ & $18.408\pm0.005$ &\\ 
BUL$\_$SC18 & 579405  & 1999-BUL-46 & 18\uph07\upm21\zdot\ups01 & $-27\arcd40\arcm04\zdot\arcs0$ &          --         &        --       &       --        & $16.648\pm0.003$ &\\ 
BUL$\_$SC18 & 594723  & 1998-BUL-15 & 18\uph07\upm20\zdot\ups83 & $-27\arcd34\arcm10\zdot\arcs8$ & $ 944.355\pm 0.080$ & $ 17.95\pm0.42$ & $13.75\pm 0.46$ & $18.675\pm0.008$ &\\ 
BUL$\_$SC18 & 596044  &      --     & 18\uph07\upm23\zdot\ups52 & $-27\arcd36\arcm18\zdot\arcs8$ & $1364.478\pm 0.375$ & $ 17.44\pm0.87$ & $ 2.32\pm 0.06$ & $18.758\pm0.008$ &\\ 
BUL$\_$SC19 & 26606   & 1999-BUL-23 & 18\uph07\upm45\zdot\ups14 & $-27\arcd33\arcm15\zdot\arcs4$ &           --        &         --      &        --       & $17.894\pm0.024$ & BCC\\
BUL$\_$SC19 & 64524   &      --     & 18\uph07\upm36\zdot\ups39 & $-27\arcd20\arcm00\zdot\arcs1$ &           --        &         --      &        --       & $18.129\pm0.009$ & BCC\\
BUL$\_$SC19 & 614666  & 1999-BUL-31 & 18\uph08\upm32\zdot\ups71 & $-27\arcd22\arcm13\zdot\arcs5$ & $1358.489\pm 0.089$ & $  5.73\pm0.22$ & $>5.1         $ & $18.188\pm0.007$ &\\ 
BUL$\_$SC20 & 300548  &      --     & 17\uph59\upm15\zdot\ups84 & $-28\arcd55\arcm49\zdot\arcs4$ & $1075.076\pm 2.317$ & $ 47.84\pm4.75$ & $ 2.28\pm 0.10$ & $18.169\pm0.011$ &\\ 
BUL$\_$SC20 & 305095  &      --     & 17\uph59\upm07\zdot\ups74 & $-28\arcd54\arcm41\zdot\arcs4$ & $1364.827\pm 0.267$ & $ 16.15\pm0.78$ & $ 2.95\pm 0.09$ & $19.088\pm0.013$ &\\ 
BUL$\_$SC20 & 391296  &      --     & 17\uph59\upm17\zdot\ups02 & $-28\arcd29\arcm41\zdot\arcs9$ & $ 963.888\pm 0.483$ & $ 42.25\pm1.85$ & $ 5.46\pm 0.16$ & $19.270\pm0.015$ & BIN?\\
BUL$\_$SC20 & 395103  &      --     & 17\uph59\upm08\zdot\ups99 & $-28\arcd24\arcm54\zdot\arcs7$ & $ 948.582\pm 0.241$ & $ 38.22\pm0.74$ & $ 3.21\pm 0.05$ & $15.435\pm0.003$ & BIN\\
BUL$\_$SC20 & 560821  &      --     & 17\uph59\upm27\zdot\ups19 & $-28\arcd32\arcm31\zdot\arcs5$ & $1051.959\pm 0.814$ & $ 16.55\pm0.76$ & $ 1.31\pm 0.02$ & $14.656\pm0.001$ &\\ 
BUL$\_$SC20 & 708586  & 1999-BUL-25 & 17\uph59\upm41\zdot\ups14 & $-28\arcd47\arcm18\zdot\arcs4$ &           --        &         --      &        --       & $18.329\pm0.022$ & BCC\\
BUL$\_$SC21 & 388360  &      --     & 18\uph00\upm06\zdot\ups73 & $-28\arcd44\arcm00\zdot\arcs2$ & $ 957.307\pm 0.303$ & $ 15.01\pm1.32$ & $ 3.75\pm 0.52$ & $18.964\pm0.011$ &\\ 
BUL$\_$SC21 & 678389  & 1998-BUL-02 & 18\uph00\upm40\zdot\ups20 & $-29\arcd19\arcm35\zdot\arcs2$ & $ 853.219\pm 6.970$ & $109.02\pm6.69$ & $ >2.9        $ & $18.158\pm0.008$ &\\ 
BUL$\_$SC21 & 766993  & 1998-BUL-41 & 18\uph00\upm39\zdot\ups38 & $-28\arcd55\arcm14\zdot\arcs6$ & $1124.910\pm 5.469$ & $ 32.37\pm3.93$ & $ 1.39\pm 0.04$ & $15.519\pm0.002$ &\\ 
BUL$\_$SC21 & 833776  &      --     & 18\uph00\upm39\zdot\ups58 & $-28\arcd34\arcm43\zdot\arcs4$ &           --        &         --      &        --       & $17.809\pm0.030$ & BCC\\
BUL$\_$SC22 & 157685  &      --     & 17\uph56\upm21\zdot\ups64 & $-30\arcd33\arcm27\zdot\arcs9$ & $ 694.751\pm 0.116$ & $ 16.30\pm0.53$ & $ 5.15\pm 0.29$ & $19.223\pm0.009$ &\\ 
BUL$\_$SC22 & 167822  & 1998-BUL-35 & 17\uph56\upm25\zdot\ups81 & $-30\arcd26\arcm56\zdot\arcs4$ & $1059.523\pm 0.135$ & $  8.28\pm0.27$ & $10.17\pm 2.34$ & $18.612\pm0.006$ &\\ 
BUL$\_$SC22 & 258372  & 1998-BUL-13 & 17\uph56\upm35\zdot\ups30 & $-30\arcd56\arcm33\zdot\arcs4$ & $ 944.963\pm 0.180$ & $ 51.27\pm0.55$ & $ 3.10\pm 0.02$ & $17.048\pm0.002$ &\\ 
BUL$\_$SC22 & 383717  &      --     & 17\uph56\upm58\zdot\ups80 & $-31\arcd14\arcm06\zdot\arcs9$ & $ 588.105\pm 0.499$ & $ 20.17\pm1.17$ & $ 3.68\pm 0.15$ & $19.136\pm0.014$ &\\ 
BUL$\_$SC22 & 390560  & 1998-BUL-34 & 17\uph56\upm56\zdot\ups81 & $-31\arcd11\arcm51\zdot\arcs0$ & $1074.195\pm 0.141$ & $ 16.75\pm0.26$ & $ 1.96\pm 0.01$ & $16.401\pm0.001$ &\\ 
BUL$\_$SC22 & 392070  &      --     & 17\uph56\upm52\zdot\ups99 & $-31\arcd10\arcm56\zdot\arcs0$ & $1042.766\pm 0.617$ & $ 10.00\pm1.20$ & $ 1.59\pm 0.04$ & $17.957\pm0.006$ &\\ 
BUL$\_$SC22 & 638508  & 1999-BUL-15 & 17\uph57\upm08\zdot\ups43 & $-30\arcd48\arcm30\zdot\arcs3$ & $1309.060\pm 0.115$ & $ 15.46\pm0.46$ & $ 5.82\pm 0.25$ & $19.335\pm0.009$ &\\ 
BUL$\_$SC23 & 240146  & 1999-BUL-04 & 17\uph57\upm54\zdot\ups05 & $-31\arcd29\arcm56\zdot\arcs6$ & $1260.582\pm 1.352$ & $ 24.17\pm1.51$ & $ 1.31\pm 0.02$ & $17.044\pm0.002$ &\\ 
BUL$\_$SC23 & 255351  &      --     & 17\uph57\upm40\zdot\ups27 & $-31\arcd24\arcm26\zdot\arcs9$ & $ 635.702\pm 0.558$ & $  9.34\pm0.69$ & $ 3.02\pm 0.40$ & $18.983\pm0.010$ &\\ 
BUL$\_$SC23 & 282632  &      --     & 17\uph57\upm46\zdot\ups55 & $-31\arcd14\arcm28\zdot\arcs3$ & $ 630.497\pm 0.147$ & $  4.24\pm0.29$ & $ 1.80\pm 0.04$ & $17.783\pm0.004$ & BIN?\\
BUL$\_$SC23 & 294092  & 1999-BUL-10 & 17\uph57\upm51\zdot\ups21 & $-31\arcd10\arcm44\zdot\arcs6$ & $1294.953\pm 0.299$ & $ 17.04\pm0.52$ & $ 2.20\pm 0.03$ & $17.925\pm0.003$ &\\ 
BUL$\_$SC23 & 340296  &      --     & 17\uph57\upm41\zdot\ups36 & $-30\arcd55\arcm33\zdot\arcs5$ & $ 559.210\pm 0.194$ & $ 31.58\pm0.50$ & $ 2.68\pm 0.02$ & $17.135\pm0.002$ &\\ 
BUL$\_$SC23 & 524386  & 1999-BUL-33 & 17\uph57\upm56\zdot\ups24 & $-30\arcd51\arcm58\zdot\arcs9$ & $1434.522\pm 0.175$ & $ 53.82\pm0.45$ & $ 2.94\pm 0.02$ & $16.669\pm0.002$ &\\ 
BUL$\_$SC24 & 57233   &      --     & 17\uph52\upm53\zdot\ups01 & $-32\arcd59\arcm02\zdot\arcs3$ & $1013.504\pm 2.139$ & $ 37.69\pm2.42$ & $ 1.17\pm 0.01$ & $16.509\pm0.002$ &\\ 
BUL$\_$SC24 & 111303  &      --     & 17\uph52\upm56\zdot\ups52 & $-32\arcd38\arcm51\zdot\arcs5$ &          --         &        --       &       --        & $18.142\pm0.006$ &\\ 
BUL$\_$SC24 & 137412  &      --     & 17\uph52\upm54\zdot\ups70 & $-32\arcd34\arcm44\zdot\arcs4$ & $1068.380\pm 1.930$ & $ 60.40\pm5.37$ & $ 2.39\pm 0.10$ & $19.376\pm0.011$ &\\ 
BUL$\_$SC24 & 198910  & 1999-BUL-08 & 17\uph53\upm02\zdot\ups78 & $-33\arcd07\arcm33\zdot\arcs5$ & $1287.493\pm 0.163$ & $ 23.53\pm0.59$ & $ 8.98\pm 0.65$ & $19.055\pm0.010$ &\\ 
BUL$\_$SC24 & 201397  &      --     & 17\uph53\upm08\zdot\ups04 & $-33\arcd06\arcm37\zdot\arcs1$ & $ 570.349\pm 0.400$ & $  7.70\pm0.46$ & $ 1.32\pm 0.02$ & $16.907\pm0.003$ &\\ 
BUL$\_$SC24 & 361426  &      --     & 17\uph53\upm19\zdot\ups43 & $-33\arcd02\arcm04\zdot\arcs8$ & $1067.635\pm 0.402$ & $ 16.04\pm0.55$ & $ 2.06\pm 0.04$ & $17.147\pm0.002$ &\\ 
BUL$\_$SC24 & 475158  &      --     & 17\uph53\upm51\zdot\ups05 & $-33\arcd14\arcm05\zdot\arcs5$ & $ 609.272\pm 0.253$ & $ 10.08\pm0.50$ & $ 2.87\pm 0.08$ & $18.514\pm0.008$ &\\ 
BUL$\_$SC25 & 610720  &      --     & 17\uph54\upm48\zdot\ups97 & $-32\arcd29\arcm23\zdot\arcs0$ & $ 910.898\pm 1.027$ & $ 50.24\pm2.88$ & $ 2.67\pm 0.09$ & $19.361\pm0.011$ &\\ 
BUL$\_$SC26 & 85908   &      --     & 17\uph46\upm48\zdot\ups16 & $-34\arcd59\arcm47\zdot\arcs3$ & $1401.554\pm 1.070$ & $ 74.64\pm2.96$ & $ 2.88\pm 0.05$ & $18.966\pm0.013$ &\\ 
BUL$\_$SC26 & 364331  &      --     & 17\uph47\upm11\zdot\ups63 & $-34\arcd34\arcm12\zdot\arcs7$ & $ 755.525\pm 5.934$ & $ 51.82\pm6.65$ & $ 4.79\pm 1.78$ & $17.967\pm0.008$ &\\ 
BUL$\_$SC26 & 395449  & 1998-BUL-09 & 17\uph47\upm23\zdot\ups43 & $-35\arcd17\arcm40\zdot\arcs9$ & $ 926.573\pm 0.835$ & $ 32.31\pm1.00$ & $ 1.61\pm 0.02$ & $17.310\pm0.002$ &\\ 
BUL$\_$SC26 & 502486  & 1999-BUL-05 & 17\uph47\upm24\zdot\ups05 & $-34\arcd46\arcm21\zdot\arcs5$ & $1274.404\pm 0.210$ & $ 33.90\pm1.28$ & $11.29\pm 0.59$ & $18.843\pm0.018$ &\\ 
BUL$\_$SC26 & 719864  &      --     & 17\uph47\upm35\zdot\ups84 & $-34\arcd31\arcm43\zdot\arcs3$ & $1087.532\pm 0.406$ & $ 11.74\pm0.46$ & $ 1.57\pm 0.01$ & $14.925\pm0.001$ &\\
\hline}

\setcounter{table}{1}
\MakeTableSep{
c@{\hspace{4pt}}
r@{\hspace{4pt}}
c@{\hspace{4pt}}
c@{\hspace{4pt}}
c@{\hspace{4pt}}
r@{\hspace{3pt}}
r@{\hspace{3pt}}
r@{\hspace{4pt}}
r@{\hspace{4pt}}
l}{12.5cm}{Continued}
{\hline
\noalign{\vskip3pt}
\multicolumn{2}{c}{Identification} & EWS & RA(2000) & DEC(2000) & 
\multicolumn{1}{c}{$T_{\rm max}$} & 
\multicolumn{1}{c}{$t_0$} & 
\multicolumn{1}{c}{$A_{\rm max}$} & \multicolumn{1}{c}{$I_0$} & Rem.\\
Field & \multicolumn{1}{c}{No} &&&&
\multicolumn{1}{c}{[HJD--2450000]} & \multicolumn{1}{c}{[days]} & &
\multicolumn{1}{c}{[mag]} &\\
\noalign{\vskip3pt}
\hline
\noalign{\vskip3pt}
BUL$\_$SC27 &  23935  &      --     & 17\uph47\upm57\zdot\ups59 & $-35\arcd27\arcm20\zdot\arcs5$ & $ 891.728\pm 1.848$ & $ 14.49\pm2.21$ & $ 1.33\pm 0.04$ & $16.747\pm0.003$ &\\ 
BUL$\_$SC27 & 331680  &      --     & 17\uph48\upm15\zdot\ups92 & $-34\arcd50\arcm08\zdot\arcs4$ & $1244.131\pm 0.325$ & $ 17.65\pm0.85$ & $ 4.98\pm 0.25$ & $18.377\pm0.011$ &\\ 
BUL$\_$SC27 & 403837  &      --     & 17\uph48\upm36\zdot\ups58 & $-35\arcd22\arcm58\zdot\arcs9$ & $ 618.973\pm 0.327$ & $ 16.02\pm0.71$ & $ 2.42\pm 0.10$ & $18.619\pm0.007$ &\\ 
BUL$\_$SC27 & 406020  &      --     & 17\uph48\upm36\zdot\ups51 & $-35\arcd22\arcm58\zdot\arcs4$ & $ 618.149\pm 0.442$ & $ 20.94\pm1.12$ & $ 2.95\pm 0.13$ & $19.055\pm0.013$ &\\ 
BUL$\_$SC27 & 568529  & 1998-BUL-05 & 17\uph48\upm42\zdot\ups80 & $-35\arcd23\arcm08\zdot\arcs5$ & $ 914.141\pm 0.451$ & $ 19.77\pm0.73$ & $ 6.66\pm 0.82$ & $18.403\pm0.006$ &\\ 
BUL$\_$SC27 & 645326  &      --     & 17\uph48\upm55\zdot\ups87 & $-34\arcd56\arcm14\zdot\arcs2$ & $1064.419\pm 1.027$ & $ 64.75\pm4.60$ & $ 4.33\pm 0.31$ & $19.499\pm0.017$ &\\ 
BUL$\_$SC29 & 391055  &      --     & 17\uph48\upm35\zdot\ups25 & $-37\arcd26\arcm49\zdot\arcs3$ & $1316.241\pm 0.517$ & $ 10.00\pm1.01$ & $ 1.78\pm 0.05$ & $18.994\pm0.009$ &\\ 
BUL$\_$SC29 & 477004  &      --     & 17\uph48\upm30\zdot\ups40 & $-36\arcd45\arcm00\zdot\arcs9$ & $1343.875\pm 0.869$ & $ 56.37\pm1.26$ & $ 1.17\pm 0.00$ & $14.717\pm0.001$ &\\ 
BUL$\_$SC30 & 57488   & 1999-BUL-06 & 18\uph01\upm02\zdot\ups52 & $-29\arcd00\arcm11\zdot\arcs6$ & $1274.320\pm 0.443$ & $ 14.31\pm0.62$ & $ 1.65\pm 0.05$ & $15.124\pm0.002$ &\\ 
BUL$\_$SC30 & 145997  &      --     & 18\uph00\upm56\zdot\ups67 & $-28\arcd36\arcm35\zdot\arcs8$ & $ 682.399\pm 0.198$ & $ 18.01\pm0.40$ & $ 6.75\pm 0.52$ & $18.030\pm0.004$ &\\ 
BUL$\_$SC30 & 165305  & 1999-BUL-12 & 18\uph01\upm07\zdot\ups74 & $-28\arcd31\arcm41\zdot\arcs7$ & $1301.833\pm 0.142$ & $ 21.99\pm0.30$ & $ 2.43\pm 0.02$ & $14.638\pm0.002$ &\\ 
BUL$\_$SC30 & 236837  & 1998-BUL-03 & 18\uph01\upm18\zdot\ups12 & $-29\arcd05\arcm49\zdot\arcs9$ & $ 893.276\pm 4.728$ & $ 58.96\pm5.23$ & $ 1.70\pm 0.03$ & $17.093\pm0.003$ &\\ 
BUL$\_$SC30 & 352272  &      --     & 18\uph01\upm21\zdot\ups14 & $-28\arcd32\arcm39\zdot\arcs6$ &           --        &         --      &        --       & $15.388\pm0.005$ & BCC\\
BUL$\_$SC30 & 471295  &      --     & 18\uph01\upm40\zdot\ups19 & $-28\arcd55\arcm30\zdot\arcs2$ & $ 620.528\pm 0.132$ & $ 17.45\pm0.62$ & $ 4.66\pm 0.27$ & $18.712\pm0.008$ &\\ 
BUL$\_$SC30 & 492925  &      --     & 18\uph01\upm40\zdot\ups96 & $-28\arcd46\arcm54\zdot\arcs9$ & $1339.032\pm 0.350$ & $  3.37\pm0.38$ & $ 1.31\pm 0.03$ & $17.537\pm0.004$ &\\ 
BUL$\_$SC30 & 499815  &      --     & 18\uph01\upm40\zdot\ups88 & $-28\arcd46\arcm54\zdot\arcs9$ & $1338.706\pm 0.158$ & $  6.91\pm0.36$ & $ 3.16\pm 0.10$ & $18.839\pm0.011$ &\\ 
BUL$\_$SC30 & 553231  &      --     & 18\uph01\upm38\zdot\ups89 & $-28\arcd30\arcm03\zdot\arcs9$ & $ 597.819\pm 0.407$ & $ 17.98\pm0.85$ & $ 2.76\pm 0.12$ & $18.503\pm0.009$ &\\ 
BUL$\_$SC30 & 559419  &      --     & 18\uph01\upm33\zdot\ups95 & $-28\arcd28\arcm02\zdot\arcs3$ & $ 710.477\pm 0.477$ & $ 29.74\pm0.50$ & $ 1.78\pm 0.02$ & $14.224\pm0.001$ &\\ 
BUL$\_$SC30 & 631133  &      --     & 18\uph01\upm51\zdot\ups70 & $-29\arcd00\arcm37\zdot\arcs5$ & $ 625.652\pm 0.052$ & $  7.16\pm0.31$ & $26.09\pm 1.96$ & $18.817\pm0.015$ &\\ 
BUL$\_$SC30 & 671185  &      --     & 18\uph01\upm47\zdot\ups58 & $-28\arcd49\arcm04\zdot\arcs9$ & $ 557.240\pm 1.244$ & $ 23.38\pm1.28$ & $ 1.31\pm 0.01$ & $16.134\pm0.001$ &\\ 
BUL$\_$SC31 & 24931   & 1998-BUL-11 & 18\uph02\upm00\zdot\ups24 & $-28\arcd57\arcm27\zdot\arcs8$ & $ 930.841\pm 0.200$ & $ 10.41\pm0.38$ & $ 2.64\pm 0.08$ & $17.726\pm0.003$ &\\ 
BUL$\_$SC31 & 48308   &      --     & 18\uph01\upm58\zdot\ups51 & $-28\arcd50\arcm08\zdot\arcs8$ & $ 643.745\pm 0.224$ & $ 10.11\pm0.23$ & $ 1.71\pm 0.02$ & $16.930\pm0.002$ &\\ 
BUL$\_$SC31 & 293442  & 1999-BUL-17 & 18\uph02\upm22\zdot\ups51 & $-28\arcd40\arcm43\zdot\arcs9$ &           --        &         --      &        --       & $18.686\pm0.015$ & BCC\\
BUL$\_$SC31 & 342145  & 1999-BUL-26 & 18\uph02\upm17\zdot\ups98 & $-28\arcd25\arcm28\zdot\arcs3$ & $1344.619\pm 0.495$ & $  6.26\pm0.66$ & $ 1.10\pm 0.01$ & $16.656\pm0.002$ &\\ 
BUL$\_$SC31 & 513194  & 1998-BUL-04 & 18\uph02\upm29\zdot\ups02 & $-28\arcd33\arcm12\zdot\arcs5$ & $ 913.672\pm 0.578$ & $ 18.25\pm0.74$ & $ 1.95\pm 0.09$ & $17.285\pm0.003$ &\\ 
BUL$\_$SC31 & 631037  & 1998-BUL-19 & 18\uph02\upm40\zdot\ups12 & $-28\arcd56\arcm28\zdot\arcs1$ & $ 965.938\pm 0.321$ & $ 28.02\pm0.93$ & $ 4.04\pm 0.09$ & $18.913\pm0.009$ &\\ 
BUL$\_$SC31 & 722752  &      --     & 18\uph02\upm39\zdot\ups20 & $-28\arcd29\arcm09\zdot\arcs3$ & $ 604.679\pm 0.575$ & $ 18.29\pm1.02$ & $ 2.17\pm 0.06$ & $18.504\pm0.008$ &\\ 
BUL$\_$SC32 & 206031  &      --     & 18\uph03\upm10\zdot\ups37 & $-28\arcd10\arcm11\zdot\arcs9$ & $ 934.680\pm 0.594$ & $ 26.41\pm1.44$ & $ 2.46\pm 0.05$ & $18.983\pm0.008$ &\\ 
BUL$\_$SC32 & 333270  &      --     & 18\uph03\upm21\zdot\ups74 & $-28\arcd28\arcm50\zdot\arcs4$ & $1015.855\pm 0.131$ & $ 12.68\pm0.26$ & $ 3.22\pm 0.43$ & $14.691\pm0.001$ &\\ 
BUL$\_$SC32 & 737024  &      --     & 18\uph03\upm45\zdot\ups39 & $-28\arcd27\arcm13\zdot\arcs0$ & $ 828.127\pm 6.633$ & $ 60.77\pm4.24$ & $ 1.59\pm 0.24$ & $16.448\pm0.003$ &\\ 
BUL$\_$SC33 & 85552   & 1998-BUL-07 & 18\uph05\upm02\zdot\ups33 & $-28\arcd55\arcm17\zdot\arcs9$ & $ 917.087\pm 0.656$ & $ 26.40\pm1.04$ & $ 3.48\pm 0.22$ & $17.534\pm0.006$ &\\ 
BUL$\_$SC33 & 164492  & 1999-BUL-32 & 18\uph05\upm05\zdot\ups34 & $-28\arcd34\arcm42\zdot\arcs3$ & $1366.382\pm 0.995$ & $166.30\pm6.17$ & $ 7.29\pm 0.31$ & $18.369\pm0.025$ &\\ 
BUL$\_$SC33 & 476067  & 1999-BUL-42 & 18\uph05\upm38\zdot\ups68 & $-28\arcd52\arcm11\zdot\arcs2$ &           --        &         --      &        --       & $18.436\pm0.011$ & BCC\\
BUL$\_$SC33 & 553617  &      --     & 18\uph05\upm46\zdot\ups71 & $-28\arcd25\arcm32\zdot\arcs1$ & $ 647.196\pm 0.920$ & $137.94\pm2.44$ & $ 2.04\pm 0.01$ & $15.818\pm0.003$ &\\ 
BUL$\_$SC34 & 87132   &      --     & 17\uph58\upm00\zdot\ups35 & $-29\arcd14\arcm59\zdot\arcs8$ &           --        &         --      &        --       & $18.551\pm0.006$ & BCC?\\ 
BUL$\_$SC34 & 144644  & 1998-BUL-17 & 17\uph57\upm47\zdot\ups60 & $-29\arcd03\arcm54\zdot\arcs3$ & $ 949.402\pm 0.129$ & $  7.73\pm0.18$ & $ 3.73\pm 0.20$ & $16.402\pm0.002$ &\\ 
BUL$\_$SC34 & 255914  &      --     & 17\uph58\upm06\zdot\ups58 & $-29\arcd35\arcm43\zdot\arcs7$ & $1459.582\pm 2.455$ & $ 41.09\pm2.22$ & $ 1.50\pm 0.03$ & $15.657\pm0.003$ &\\ 
BUL$\_$SC34 & 472357  &      --     & 17\uph58\upm12\zdot\ups49 & $-28\arcd44\arcm50\zdot\arcs1$ & $ 602.275\pm 0.178$ & $ 20.49\pm0.80$ & $16.91\pm 1.32$ & $19.004\pm0.010$ &\\ 
BUL$\_$SC34 & 639703  &      --     & 17\uph58\upm22\zdot\ups81 & $-28\arcd59\arcm55\zdot\arcs0$ & $ 613.829\pm 0.233$ & $ 19.70\pm0.57$ & $ 3.17\pm 0.08$ & $17.831\pm0.006$ &\\ 
BUL$\_$SC34 & 651798  & 1998-BUL-21 & 17\uph58\upm33\zdot\ups77 & $-28\arcd54\arcm28\zdot\arcs7$ & $ 992.410\pm 0.632$ & $ 25.57\pm1.29$ & $ 2.44\pm 0.06$ & $15.590\pm0.003$ &\\ 
BUL$\_$SC34 & 840343  &      --     & 17\uph58\upm37\zdot\ups12 & $-29\arcd06\arcm29\zdot\arcs9$ & $ 798.792\pm 0.414$ & $ 72.02\pm0.81$ & $ 4.39\pm 0.91$ & $13.648\pm0.001$ &\\ 
BUL$\_$SC35 & 54409   & 1998-BUL-36 & 18\uph04\upm02\zdot\ups87 & $-28\arcd09\arcm22\zdot\arcs4$ & $1083.910\pm 0.164$ & $ 12.10\pm0.57$ & $ 3.01\pm 0.16$ & $17.312\pm0.003$ &\\ 
BUL$\_$SC35 & 144974  &      --     & 18\uph04\upm09\zdot\ups65 & $-27\arcd44\arcm34\zdot\arcs9$ & $ 684.853\pm 0.266$ & $ 27.14\pm0.40$ & $ 1.87\pm 0.01$ & $15.730\pm0.001$ &\\ 
BUL$\_$SC35 & 305604  &      --     & 18\uph04\upm22\zdot\ups41 & $-27\arcd57\arcm52\zdot\arcs3$ &           --        &         --      &        --       & $19.035\pm0.020$ & BCC\\
BUL$\_$SC35 & 451130  & 1998-BUL-23 & 18\uph04\upm33\zdot\ups63 & $-28\arcd07\arcm32\zdot\arcs2$ & $ 997.439\pm 1.445$ & $ 16.05\pm1.82$ & $ 1.74\pm 0.05$ & $15.390\pm0.004$ & BLE\\
BUL$\_$SC35 & 770398  &      --     & 18\uph04\upm54\zdot\ups02 & $-27\arcd30\arcm07\zdot\arcs0$ & $1299.511\pm 1.639$ & $ 24.91\pm3.34$ & $ 2.04\pm 0.10$ & $19.017\pm0.019$ &\\ 
BUL$\_$SC36 & 88551   &      --     & 18\uph05\upm02\zdot\ups67 & $-28\arcd00\arcm47\zdot\arcs7$ &          --         &        --       &       --        & $16.878\pm0.003$ &\\ 
BUL$\_$SC36 & 336761  & 1999-BUL-11 & 18\uph05\upm29\zdot\ups59 & $-27\arcd59\arcm14\zdot\arcs4$ &           --        &         --      &        --       & $18.600\pm0.020$ & BCC\\
BUL$\_$SC37 & 92757   &      --     & 17\uph52\upm13\zdot\ups29 & $-29\arcd52\arcm25\zdot\arcs8$ & $ 689.645\pm 0.289$ & $ 14.37\pm0.72$ & $ 2.79\pm 0.07$ & $18.279\pm0.007$ &\\ 
BUL$\_$SC37 & 226174  &      --     & 17\uph52\upm20\zdot\ups66 & $-30\arcd04\arcm20\zdot\arcs5$ & $ 557.512\pm 0.049$ & $  7.13\pm0.15$ & $ 4.05\pm 0.07$ & $18.558\pm0.005$ &\\ 
BUL$\_$SC37 & 321846  &      --     & 17\uph52\upm20\zdot\ups50 & $-29\arcd31\arcm15\zdot\arcs4$ &          --         &        --       &       --        & $17.151\pm0.002$ &\\ 
BUL$\_$SC37 & 366449  &      --     & 17\uph52\upm33\zdot\ups50 & $-30\arcd13\arcm33\zdot\arcs5$ & $ 645.121\pm 0.174$ & $ 14.66\pm0.49$ & $ 5.63\pm 0.22$ & $18.442\pm0.009$ &\\ 
BUL$\_$SC37 & 369822  & 1999-BUL-20 & 17\uph52\upm36\zdot\ups84 & $-30\arcd11\arcm30\zdot\arcs6$ & $1317.037\pm 0.126$ & $  2.38\pm0.14$ & $ 1.29\pm 0.02$ & $15.289\pm0.001$ &\\ 
\hline}

\setcounter{table}{1}
\MakeTableSep{
c@{\hspace{4pt}}
r@{\hspace{4pt}}
c@{\hspace{4pt}}
c@{\hspace{4pt}}
c@{\hspace{4pt}}
r@{\hspace{3pt}}
r@{\hspace{3pt}}
r@{\hspace{4pt}}
r@{\hspace{4pt}}
l}{12.5cm}{Concluded}
{\hline
\noalign{\vskip3pt}
\multicolumn{2}{c}{Identification} & EWS & RA(2000) & DEC(2000) & 
\multicolumn{1}{c}{$T_{\rm max}$} & 
\multicolumn{1}{c}{$t_0$} & 
\multicolumn{1}{c}{$A_{\rm max}$} & \multicolumn{1}{c}{$I_0$} & Rem.\\
Field & \multicolumn{1}{c}{No} &&&&
\multicolumn{1}{c}{[HJD--2450000]} & \multicolumn{1}{c}{[days]} & &
\multicolumn{1}{c}{[mag]} &\\
\noalign{\vskip3pt}
\hline
\noalign{\vskip3pt}
BUL$\_$SC37 & 552723  &      --     & 17\uph52\upm59\zdot\ups18 & $-30\arcd06\arcm20\zdot\arcs7$ & $ 628.811\pm 0.154$ & $  6.65\pm0.37$ & $ 2.31\pm 0.05$ & $18.608\pm0.008$ &\\ 
BUL$\_$SC37 & 589087  &      --     & 17\uph52\upm56\zdot\ups46 & $-29\arcd54\arcm00\zdot\arcs0$ & $ 690.985\pm 0.603$ & $ 40.31\pm1.00$ & $ 1.87\pm 0.02$ & $17.554\pm0.003$ &\\ 
BUL$\_$SC37 & 655063  & 1999-BUL-34 & 17\uph52\upm48\zdot\ups61 & $-29\arcd32\arcm35\zdot\arcs1$ & $1369.649\pm 0.055$ & $  6.09\pm0.15$ & $ 2.69\pm 0.02$ & $16.441\pm0.002$ &\\ 
BUL$\_$SC38 & 95103   & 1998-BUL-22 & 18\uph01\upm09\zdot\ups74 & $-29\arcd56\arcm18\zdot\arcs9$ & $ 990.465\pm 0.028$ & $  6.97\pm0.10$ & $ 6.48\pm 0.17$ & $16.019\pm0.002$ &\\ 
BUL$\_$SC38 & 120518  & 1999-BUL-07 & 18\uph01\upm10\zdot\ups23 & $-29\arcd48\arcm55\zdot\arcs2$ & $1316.252\pm 0.155$ & $ 34.36\pm0.28$ & $ 2.09\pm 0.01$ & $15.973\pm0.001$ &\\ 
BUL$\_$SC38 & 447654  &      --     & 18\uph01\upm34\zdot\ups43 & $-30\arcd00\arcm07\zdot\arcs7$ &           --        &         --      &        --       & $17.405\pm0.003$ &\\ 
BUL$\_$SC38 & 589586  & 1999-BUL-18 & 18\uph01\upm47\zdot\ups98 & $-30\arcd09\arcm37\zdot\arcs7$ & $1319.863\pm 0.972$ & $ 26.71\pm1.53$ & $ 1.74\pm 0.04$ & $18.379\pm0.008$ &\\ 
BUL$\_$SC38 & 627315  & 1999-BUL-37 & 18\uph01\upm51\zdot\ups02 & $-29\arcd56\arcm50\zdot\arcs1$ & $1399.084\pm 1.217$ & $ 22.45\pm1.09$ & $ 1.31\pm 0.02$ & $16.254\pm0.002$ &\\ 
BUL$\_$SC39 & 259656  & 1999-BUL-45 & 17\uph55\upm34\zdot\ups31 & $-30\arcd01\arcm06\zdot\arcs6$ & $1420.352\pm 0.894$ & $ 27.99\pm1.42$ & $ 1.53\pm 0.03$ & $17.757\pm0.003$ & DLS \\ 
BUL$\_$SC39 & 260832  &      --     & 17\uph55\upm23\zdot\ups48 & $-30\arcd02\arcm09\zdot\arcs5$ & $ 652.696\pm 0.512$ & $ 14.53\pm0.87$ & $ 1.97\pm 0.05$ & $18.202\pm0.007$ &\\ 
BUL$\_$SC39 & 322789  &      --     & 17\uph55\upm26\zdot\ups87 & $-29\arcd43\arcm59\zdot\arcs7$ & $ 968.320\pm 1.267$ & $ 23.77\pm1.23$ & $ 1.13\pm 0.01$ & $15.706\pm0.001$ &\\ 
BUL$\_$SC39 & 323517  &      --     & 17\uph55\upm36\zdot\ups42 & $-29\arcd42\arcm14\zdot\arcs0$ & $ 703.221\pm 0.298$ & $ 15.46\pm0.40$ & $ 1.46\pm 0.01$ & $16.168\pm0.001$ &\\ 
BUL$\_$SC39 & 361372  &      --     & 17\uph55\upm28\zdot\ups67 & $-29\arcd33\arcm41\zdot\arcs7$ & $ 682.308\pm 0.053$ & $ 18.18\pm0.09$ & $ 5.55\pm 0.11$ & $13.392\pm0.001$ &\\ 
BUL$\_$SC39 & 362291  &      --     & 17\uph55\upm29\zdot\ups29 & $-29\arcd32\arcm08\zdot\arcs8$ & $1244.458\pm 2.484$ & $ 61.40\pm2.13$ & $ 1.18\pm 0.00$ & $15.679\pm0.001$ &\\ 
BUL$\_$SC39 & 378193  &      --     & 17\uph55\upm30\zdot\ups01 & $-29\arcd27\arcm38\zdot\arcs6$ & $ 588.726\pm 0.417$ & $ 13.49\pm0.71$ & $ 2.44\pm 0.07$ & $17.917\pm0.010$ & BIN \\ 
BUL$\_$SC39 & 442460  &      --     & 17\uph55\upm48\zdot\ups72 & $-30\arcd07\arcm09\zdot\arcs6$ &           --        &         --      &        --       & $19.353\pm0.013$ &\\ 
BUL$\_$SC39 & 448437  & 1998-BUL-38 & 17\uph55\upm51\zdot\ups91 & $-30\arcd02\arcm31\zdot\arcs4$ & $1092.067\pm 0.398$ & $ 13.46\pm0.39$ & $ 1.42\pm 0.01$ & $14.746\pm0.001$ &\\ 
BUL$\_$SC39 & 468687  &      --     & 17\uph55\upm45\zdot\ups16 & $-29\arcd56\arcm38\zdot\arcs6$ & $1207.733\pm 4.491$ & $ 50.65\pm2.46$ & $ 1.11\pm 0.01$ & $15.885\pm0.001$ &\\ 
BUL$\_$SC39 & 505074  &      --     & 17\uph55\upm53\zdot\ups59 & $-29\arcd47\arcm55\zdot\arcs3$ & $1222.921\pm 3.314$ & $ 14.97\pm2.83$ & $ 2.00\pm 0.30$ & $17.907\pm0.004$ &\\ 
BUL$\_$SC39 & 576039  &      --     & 17\uph55\upm43\zdot\ups97 & $-29\arcd24\arcm28\zdot\arcs7$ & $ 623.538\pm 1.563$ & $ 34.07\pm1.52$ & $ 1.09\pm 0.00$ & $16.076\pm0.001$ &\\ 
BUL$\_$SC39 & 623258  & 1999-BUL-01 & 17\uph56\upm04\zdot\ups05 & $-30\arcd08\arcm53\zdot\arcs0$ & $1171.451\pm 1.574$ & $ 90.89\pm2.78$ & $ 1.45\pm 0.03$ & $17.016\pm0.002$ &\\ 
BUL$\_$SC39 & 702483  &      --     & 17\uph56\upm01\zdot\ups61 & $-29\arcd41\arcm41\zdot\arcs8$ & $ 696.073\pm 0.089$ & $ 16.21\pm0.41$ & $ 7.85\pm 0.39$ & $18.239\pm0.006$ &\\ 
BUL$\_$SC39 & 729484  &      --     & 17\uph55\upm56\zdot\ups72 & $-29\arcd35\arcm01\zdot\arcs0$ & $ 650.882\pm 1.941$ & $138.87\pm7.81$ & $ 2.68\pm 0.08$ & $19.497\pm0.026$ &\\ 
BUL$\_$SC40 & 40759   &      --     & 17\uph50\upm43\zdot\ups80 & $-33\arcd27\arcm47\zdot\arcs6$ & $1309.017\pm 0.301$ & $  3.73\pm0.40$ & $ 1.31\pm 0.04$ & $16.868\pm0.003$ &\\ 
BUL$\_$SC40 & 415182  &      --     & 17\uph51\upm14\zdot\ups44 & $-33\arcd10\arcm30\zdot\arcs5$ & $ 974.833\pm 1.662$ & $ 11.61\pm2.00$ & $ >4.3        $ & $18.177\pm0.011$ &\\ 
BUL$\_$SC40 & 434222  & 1999-BUL-19 & 17\uph51\upm10\zdot\ups76 & $-33\arcd03\arcm44\zdot\arcs1$ & $1324.063\pm 2.443$ & $127.35\pm5.16$ & $ 1.31\pm 0.00$ & $16.072\pm0.002$ & DLS \\
BUL$\_$SC40 & 497858  & 1999-BUL-22 & 17\uph51\upm31\zdot\ups82 & $-33\arcd34\arcm00\zdot\arcs9$ & $1323.572\pm 0.177$ & $  7.05\pm0.27$ & $ 3.00\pm 0.18$ & $17.715\pm0.006$ &\\ 
BUL$\_$SC40 & 573818  &      --     & 17\uph51\upm31\zdot\ups79 & $-33\arcd04\arcm42\zdot\arcs0$ &          --         &        --       &       --        & $15.261\pm0.001$ &\\ 
BUL$\_$SC40 & 574713  &      --     & 17\uph51\upm31\zdot\ups66 & $-33\arcd04\arcm51\zdot\arcs0$ & $ 551.049\pm 0.495$ & $  6.94\pm0.57$ & $ 2.54\pm 0.15$ & $17.351\pm0.003$ &\\ 
BUL$\_$SC41 & 100122  &      --     & 17\uph51\upm42\zdot\ups97 & $-32\arcd58\arcm19\zdot\arcs2$ & $1059.693\pm 3.174$ & $ 45.29\pm3.90$ & $ 1.24\pm 0.02$ & $17.467\pm0.003$ &\\ 
BUL$\_$SC41 & 172497  &      --     & 17\uph52\upm00\zdot\ups69 & $-33\arcd26\arcm36\zdot\arcs3$ & $ 683.150\pm 1.492$ & $ 27.57\pm1.79$ & $ 1.16\pm 0.01$ & $15.939\pm0.001$ &\\ 
BUL$\_$SC41 & 345663  &      --     & 17\uph52\upm14\zdot\ups52 & $-33\arcd21\arcm39\zdot\arcs1$ & $ 888.851\pm 1.800$ & $ 71.77\pm4.54$ & $ 3.38\pm 0.10$ & $18.338\pm0.012$ &\\ 
BUL$\_$SC41 & 353302  &      --     & 17\uph52\upm23\zdot\ups57 & $-33\arcd15\arcm39\zdot\arcs0$ &          --         &        --       &       --        & $15.329\pm0.001$ &\\ 
BUL$\_$SC41 & 392439  &      --     & 17\uph52\upm16\zdot\ups01 & $-33\arcd02\arcm49\zdot\arcs1$ & $ 982.483\pm 0.716$ & $ 10.38\pm0.88$ & $ 1.77\pm 0.38$ & $17.429\pm0.003$ &\\ 
BUL$\_$SC41 & 402647  &      --     & 17\uph52\upm17\zdot\ups02 & $-33\arcd00\arcm06\zdot\arcs5$ & $ 570.259\pm 0.920$ & $ 29.66\pm2.43$ & $ 2.11\pm 0.06$ & $18.858\pm0.014$ &\\ 
BUL$\_$SC41 & 428823  & 1998-BUL-10 & 17\uph52\upm19\zdot\ups07 & $-32\arcd48\arcm20\zdot\arcs8$ & $ 927.411\pm 0.477$ & $ 51.32\pm1.45$ & $ 3.94\pm 0.08$ & $18.998\pm0.009$ &\\ 
BUL$\_$SC41 & 568646  & 1998-BUL-40 & 17\uph52\upm31\zdot\ups53 & $-32\arcd50\arcm53\zdot\arcs0$ & $1127.438\pm 7.665$ & $ 65.34\pm7.71$ & $ 2.07\pm 0.21$ & $17.979\pm0.010$ &\\ 
BUL$\_$SC42 & 262259  & 1999-BUL-28 & 18\uph09\upm04\zdot\ups93 & $-26\arcd44\arcm38\zdot\arcs3$ & $1346.786\pm 0.090$ & $  5.97\pm0.22$ & $ 3.26\pm 0.07$ & $17.924\pm0.005$ &\\ 
BUL$\_$SC42 & 347998  & 1998-BUL-32 & 18\uph09\upm10\zdot\ups54 & $-27\arcd06\arcm42\zdot\arcs0$ & $1028.081\pm 0.222$ & $ 33.09\pm0.95$ & $ >23.4       $ & $18.332\pm0.014$ &\\ 
BUL$\_$SC42 & 402431  &      --     & 18\uph09\upm20\zdot\ups54 & $-26\arcd45\arcm47\zdot\arcs9$ & $1058.603\pm 3.930$ & $ 54.12\pm5.05$ & $ 1.33\pm 0.02$ & $17.597\pm0.004$ &\\ 
BUL$\_$SC42 & 591554  & 1999-BUL-43 & 18\uph09\upm29\zdot\ups58 & $-26\arcd29\arcm48\zdot\arcs7$ & $1405.681\pm 0.468$ & $ 13.77\pm1.33$ & $ 2.57\pm 0.16$ & $18.694\pm0.011$ &\\ 
BUL$\_$SC43 & 26773   &      --     & 17\uph34\upm52\zdot\ups66 & $-27\arcd23\arcm14\zdot\arcs4$ & $1054.453\pm 0.609$ & $ 51.71\pm0.94$ & $ 1.70\pm 0.01$ & $17.184\pm0.002$ &\\ 
BUL$\_$SC43 & 130624  &      --     & 17\uph35\upm07\zdot\ups50 & $-27\arcd30\arcm37\zdot\arcs9$ & $1068.121\pm 0.162$ & $ 15.22\pm0.25$ & $ 2.51\pm 0.08$ & $17.173\pm0.001$ &\\ 
BUL$\_$SC43 & 243401  & 1999-BUL-16 & 17\uph35\upm13\zdot\ups80 & $-27\arcd34\arcm38\zdot\arcs8$ & $1334.421\pm 0.703$ & $ 25.05\pm1.13$ & $ 1.50\pm 0.02$ & $17.000\pm0.004$ &\\ 
BUL$\_$SC43 & 319475  &      --     & 17\uph35\upm24\zdot\ups55 & $-26\arcd59\arcm28\zdot\arcs2$ & $ 977.055\pm 0.412$ & $  7.82\pm0.50$ & $ 1.51\pm 0.04$ & $17.177\pm0.002$ &\\ 
BUL$\_$SC44 & 17823   & 1999-BUL-35 & 17\uph49\upm01\zdot\ups66 & $-30\arcd15\arcm40\zdot\arcs4$ & $1391.811\pm 0.185$ & $ 28.96\pm0.52$ & $ >10.3       $ & $18.932\pm0.008$ &\\ 
BUL$\_$SC44 & 20145   &      --     & 17\uph48\upm56\zdot\ups55 & $-30\arcd11\arcm40\zdot\arcs7$ & $ 606.604\pm 0.995$ & $100.31\pm2.93$ & $ 1.76\pm 0.01$ & $15.876\pm0.003$ &\\ 
BUL$\_$SC44 & 151528  &      --     & 17\uph49\upm24\zdot\ups51 & $-30\arcd24\arcm37\zdot\arcs5$ & $1009.399\pm 1.404$ & $ 68.43\pm3.72$ & $ 2.11\pm 0.06$ & $19.159\pm0.011$ &\\ 
BUL$\_$SC44 & 172247  &      --     & 17\uph49\upm33\zdot\ups13 & $-30\arcd11\arcm56\zdot\arcs1$ & $1359.698\pm 3.875$ & $ 55.78\pm5.51$ & $ 1.24\pm 0.02$ & $17.984\pm0.006$ &\\ 
BUL$\_$SC44 & 222903  & 1998-BUL-29 & 17\uph49\upm22\zdot\ups49 & $-29\arcd36\arcm52\zdot\arcs9$ & $1042.270\pm 0.007$ & $ 14.67\pm0.15$ & $ >46.6       $ & $19.267\pm0.008$ &\\ 
BUL$\_$SC44 & 239547  &      --     & 17\uph49\upm44\zdot\ups37 & $-30\arcd23\arcm45\zdot\arcs5$ & $ 571.005\pm 0.499$ & $  6.58\pm0.64$ & $ 1.31\pm 0.02$ & $17.904\pm0.004$ &\\ 
BUL$\_$SC44 & 241383  &      --     & 17\uph49\upm43\zdot\ups53 & $-30\arcd21\arcm27\zdot\arcs6$ & $1242.977\pm 1.276$ & $ 14.34\pm1.65$ & $ 1.57\pm 0.08$ & $18.374\pm0.006$ &\\ 
BUL$\_$SC44 & 256957  & 1998-BUL-25 & 17\uph49\upm40\zdot\ups00 & $-30\arcd10\arcm12\zdot\arcs0$ & $1041.409\pm 0.376$ & $ 31.80\pm0.80$ & $ 2.03\pm 0.02$ & $17.760\pm0.003$ &\\ 
BUL$\_$SC44 & 288419  & 1998-BUL-20 & 17\uph49\upm49\zdot\ups15 & $-29\arcd49\arcm10\zdot\arcs3$ & $ 967.183\pm 0.160$ & $  9.48\pm0.32$ & $ 1.75\pm 0.01$ & $16.863\pm0.002$ &\\ 
BUL$\_$SC44 & 295463  &      --     & 17\uph49\upm52\zdot\ups76 & $-29\arcd45\arcm27\zdot\arcs1$ & $ 934.438\pm 0.312$ & $ 15.24\pm0.66$ & $ 2.44\pm 0.04$ & $17.858\pm0.004$ &\\ 
\hline
\multicolumn{10}{l}{BCC -- caustic crossing binary microlensing, BIN -- possible binary microlensing}\\
\multicolumn{10}{l}{DLS -- binary microlensing or double source star, BLE -- severely
blended, VAR -- possible variable star.}\\
}

\Section{Catalog of Microlensing Events}

The Catalog of Microlensing Events in the Galactic Bulge consists of
two parts: the table with basic data of the microlensing event and
coordinates of the lensed object and the atlas presenting photometry of
the event.

Table~2 lists all microlensing event candidates presented in the
Catalog. The first column contains the ID of the lensed object in
regular OGLE convention \ie consisting of the field name and database
number. In the next column we provide cross identification with the EWS
name when the microlensing event was detected in real time. The two next
columns contain equatorial coordinates (J2000) of the lensed star
followed by parameters of the microlensing event. The latter were
calculated assuming point mass microlensing and include the moment of
maximum brightness (HJD--2450000), $T_0$, time of the Einstein radius
crossing, $t_0$, magnification at maximum brightness, $A_{\rm max}$, and
{\it I}-band baseline magnitude. The final column of Table~2 includes
remarks on non-standard events. In particular we note there candidates
for binary microlensing. When such a candidate is a typical caustic
crossing binary event we do not provide single point mass microlensing
fit parameters because they are meaningless in such cases and modeling
of light curve requires much more complicated approach. In weak cases of
possible binary microlensing single mass parameters are provided.
Abbreviations in the remark column have the following meaning: {\sc bcc}
-- caustic crossing binary microlensing; {\sc bin} -- possible binary
microlensing; {\sc dls} -- binary microlensing or double source star;
{\sc ble} -- severely blended event; {\sc var} -- possible variable
star.

The atlas of light curves of microlensing events in the Galactic bulge
detected during the seasons 1997-1999 is presented in Appendix. We do
not provide there finding charts, but they are available electronically
from the OGLE Internet archive. Also photometric data of all presented
events can be retrieved from there.

\Section{Discussion}

Two hundred fourteen microlensing event candidates were detected in the
OGLE-II fields during the observing seasons 1997--1999. Significant
number of events from 1998--1999 seasons, \ie when the EWS alert system
was implemented, were detected in real time. Many of these events were
then followed up by other groups for detailed study (\eg 1998-BUL-14,
Albrow \etal 2000a).  Because part of the OGLE-II fields overlap with
those observed by the MACHO team several events were discovered by both
teams. Cross-identification of those discovered by alert systems can be
found on the WWW alert pages of both teams while for the remaining
objects (in particular all 1997 season events) it can be done using
accurate equatorial coordinates provided in this paper and on the MACHO
WWW alert page.

The sample of OGLE-II microlensing events contains several cases of
evident lensing caused by binary object. The sample of characteristic
caustic crossing events consists of 14 events. In five additional cases
it is possible that the observed light variations are also caused by
binary lens. In two cases, BUL$\_$SC39 259656 and BUL$\_$SC40 434222,
two separate microlensing like episodes were detected. These cases might
be explained either as due to  binary microlensing without caustic
crossing or double lensed star. In the latter case the time scales of
both episodes should be equal what can discriminate between those two
possibilities. Time scales of BUL$\_$SC39 259656 episodes are equal to
about 15 and 27~days suggesting that this case is a binary microlensing.
The second brightening episode of BUL$\_$SC40 434222 started at the end
of 1999 season but observations collected up to the moment of writing
this paper confirm that it can be of microlensing origin (see OGLE EWS
WWW page). In this case the time scales are within errors the same
($t_0\approx 130$~days) suggesting double source star. However, the
second episode is still in the rising part of its light curve.

\begin{figure}[htb]
\psfig{figure=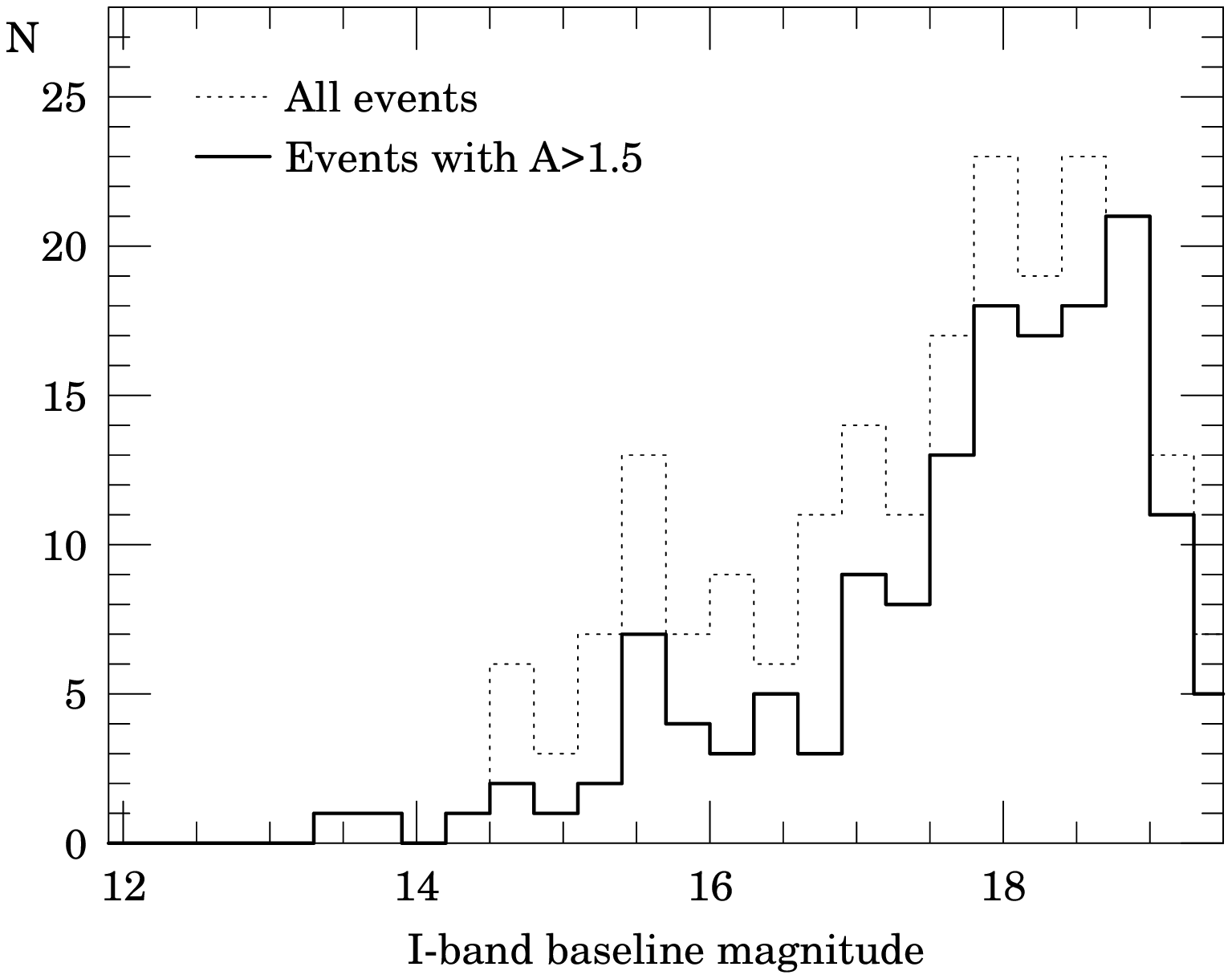,bbllx=60pt,bblly=50pt,bburx=505pt,bbury=405pt,width=12cm,clip=}
\FigCap{Distribution of the {\it I}-band baseline magnitude of lensed
stars.}
\end{figure}

The rate of the binary caustic crossing microlensing toward the Galactic
bulge observed in our sample of all microlensing events is equal to
$\approx 6.5$\% while that of all binary microlensing $\approx 9.3$\%.
It is worth noting that this is much more than reported by the MACHO
team (Alcock \etal 1999). On the other hand Mao and Paczy{\'n}ski (1991)
in their classical paper on binary microlensing predicted about 10\%
rate of strong binary microlensing events. The agreement between the
predicted and observed rates is very good.

Among many interesting cases of microlensing presented in the Catalog we
draw attention to the object, BUL$\_$SC5 244353. Unfortunately, only
falling part of its light curve was covered during the entire
observation period. While the shape of the light curve strongly
resembles that of microlensing event, it cannot be excluded that the
object is a variable star. If, however, its brightening was caused by
microlensing, the time-scale of the event was very long. The event would
closely resemble those reported by Bennett \etal (1999b) which are
supposed to be caused by black hole lenses.

\begin{figure}[htb]
\psfig{figure=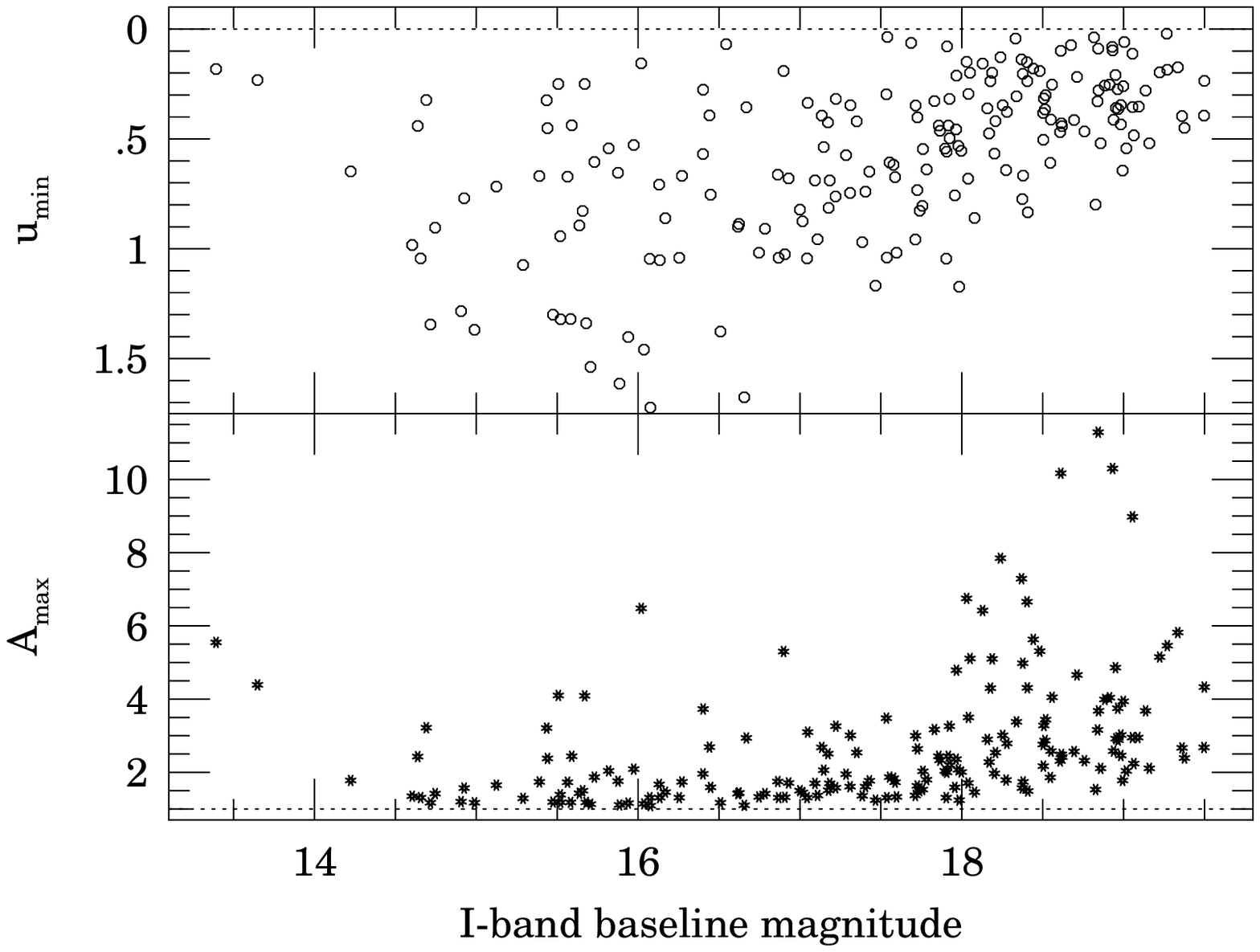,bbllx=60pt,bblly=50pt,bburx=505pt,bbury=405pt,width=12cm,clip=}
\FigCap{Magnification at maximum, $A_{\rm max}$, and minimum impact parameter,
$u_{\rm min}$, as a function of the {\it I}-band baseline magnitude.
}
\end{figure}

The statistic of collected events toward the Galactic bulge is becoming
significantly larger after each observing season. While an accurate
analysis requires additional information like  efficiency of detection,
the number of presented microlensing events is already so large  that
interesting correlations can be shown.

Efficiency of detection of microlensing events depends at least on two
factors: time sampling of the microlensing light curve and brightness of
the lensed star. Typical sampling of one/two observations per night
limits detection to events with sufficiently long time-scale. For
fainter lensed stars the $3\sigma$ detection threshold is larger,
therefore only larger magnification events can be triggered. In Fig.~2
we plot distribution of the {\it I}-band baseline magnitude of lensed
stars from our sample. The bins are 0.3~mag wide. Dotted line presents
the distribution of all events while the bold solid line the one of
subsample of events with magnification $A_{\rm max}>1.5$. The shape of
both distributions indicate that the entire sample of microlensing
events is reasonably complete up to  $I\approx18.0$~mag while subsample
of events with $A_{\rm max}>1.5$ up to $I\approx18.8$~mag.

Selection effect due to brightness of the lensed star can also be
assessed from Fig.~3 which shows magnification at the maximum,$A_{\rm
max}$, and minimum impact parameter, $u_{\rm min}$, plotted against the
{\it I}-band baseline magnitude. It is clearly seen from Fig.~3 that our
sample is quite complete for stars of $I<18$~mag and $A_{\rm max}>1.3$,
\ie $u<1.0$. As can be expected for fainter stars only higher
magnification events were triggered. Nevertheless at $I=19$~mag the
limit of reasonable completeness is still $A_{\rm max}\approx1.5$ 
($u_{\rm min}\approx0.8$).

\begin{figure}[htb]
\psfig{figure=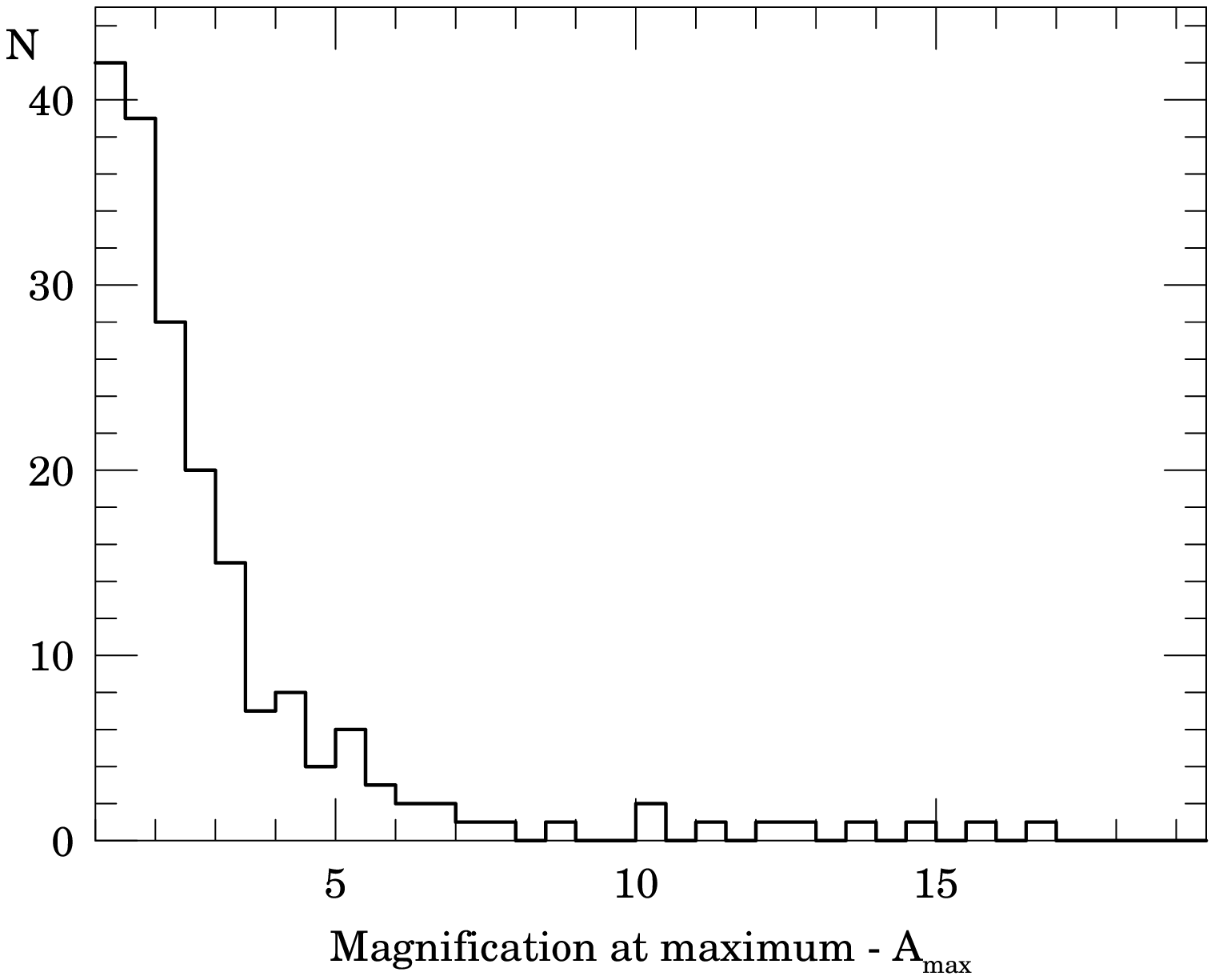,bbllx=60pt,bblly=50pt,bburx=505pt,bbury=405pt,width=12.5cm,clip=}
\FigCap{Distribution of the maximum magnification $A_{\rm max}$.}
\end{figure}

Fig.~4 presents distribution of maximum magnification of our entire
sample. This parameter ranges from as low as 1.1 up to about 50 (in a
few cases the single point mass model predicts extremely large
magnifications but because of lack of observations at the very maximum
these values are not reliable.  In these cases we provide lower limit of
magnification resulting from the brightest observation). Below $A_{\rm
max}=1.3$ incompleteness of our sample is larger -- only for brighter
stars so low magnification events could be detected. However, the
completeness becomes much higher for events with $A_{\rm max}>1.5$. The
number of events is gradually falling from $A_{\rm max}\approx1.7$ to
$A_{\rm max}=7$ with a long tail of single events with higher $A_{\rm
max}$.

\begin{figure}[htb]
\psfig{figure=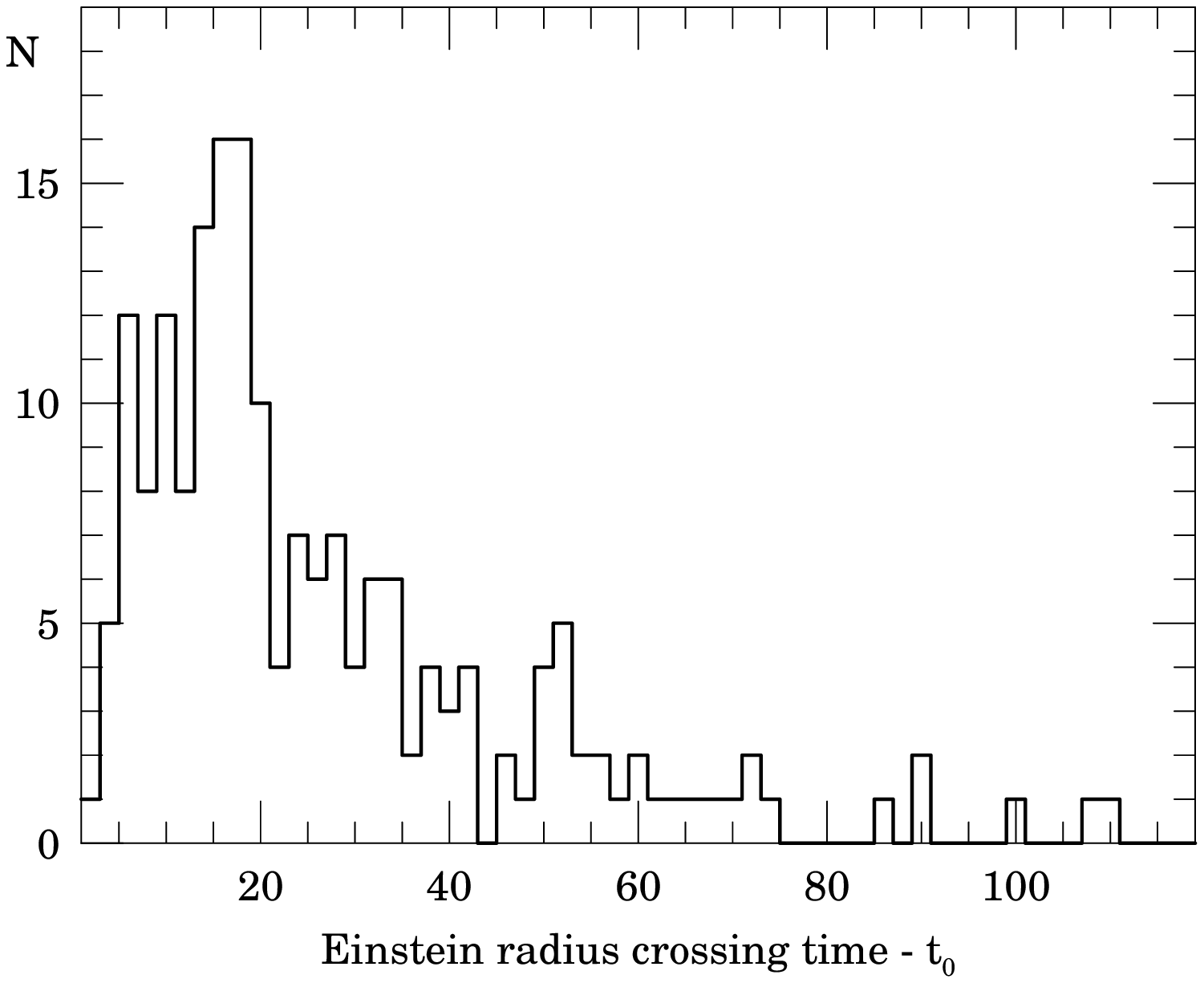,bbllx=60pt,bblly=50pt,bburx=505pt,bbury=405pt,width=12.5cm,clip=}
\FigCap{Distribution of the Einstein radius crossing time $t_0$.}
\end{figure}

Fig.~5 shows the distribution of the Einstein radius crossing time,
$t_0$, for our sample of microlensing events. This parameter is also
likely to be affected by incompleteness resulting from the sampling of
the light curve. Although we have not performed yet detailed analysis of
the dependence of detection efficiency on the event time-scale  for our
entire data set, preliminary tests performed on 1997 databases of
constant stars in similar manner as in Udalski \etal (1994c) indicate
that the region of reasonable efficiency is extended toward shorter time
scales  as compared to the OGLE-I phase. For events with $t_0>8$ days
efficiency of detection becomes relatively flat, so we may expect that
the distribution of $t_0$ is also relatively complete for events longer
than that limit. The distribution of $t_0$ peaks at $t_0\approx17$~days
with a long tail of longer time-scale events. One should also note  a
small excess of events with $t_0\approx 50$~days. It was marginally seen
in the MACHO data (Alcock \etal 1997a). The time-scale is plotted as a
function of the {\it I}-band baseline magnitude in Fig.~6. No evident
correlation is seen in this plot.

\begin{figure}[htb]
\psfig{figure=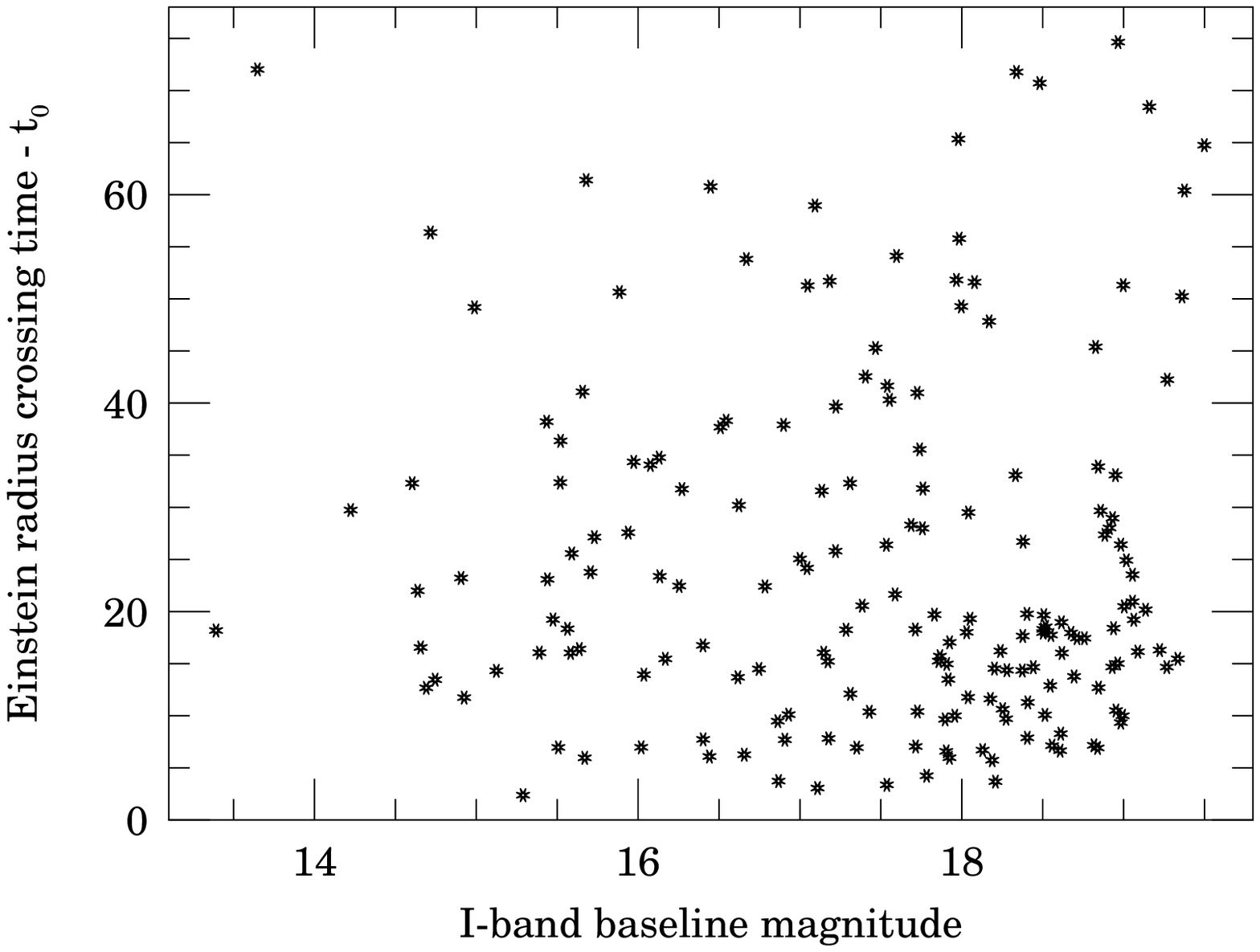,bbllx=60pt,bblly=50pt,bburx=505pt,bbury=405pt,width=12.5cm,clip=}
\FigCap{The Einstein radius crossing time, $t_0$, as a function of the
{\it I}-band baseline magnitude. }
\end{figure}

\begin{figure}[htb]
\psfig{figure=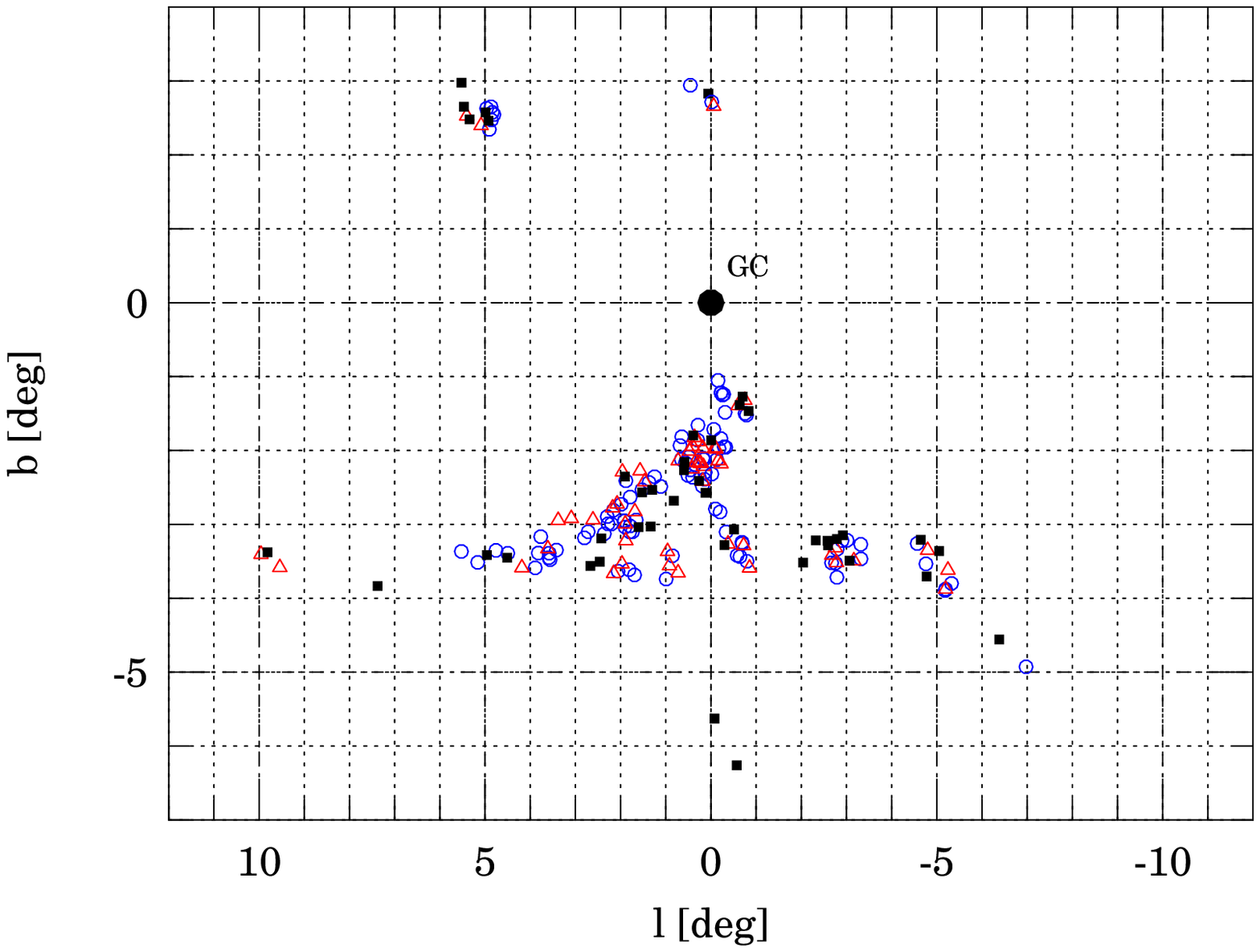,bbllx=35pt,bblly=50pt,bburx=505pt,bbury=405pt,width=12.5cm,clip=}
\FigCap{Spatial distribution in the Galactic bulge of the short
(open circles), medium (open triangles) and long (filled squares)
time-scale microlensing events.}
\end{figure}

The spatial distribution of events with different time-scales is
presented in Fig.~7. The total sample was divided into three sub-samples
of short ($t_0<20$~days), medium ($20<t_0<40$~days) and long time-scale
($t_0>40$~days) events. Location on the sky of these three samples in
the Galactic coordinates ($l,b$) is plotted with different symbols in
Fig.~7. If the different time-scale events were to be caused by
different populations of lensing objects one could expect some
differences in the distribution of our three sub-samples on the sky.
However, this does not seem to be true -- the distribution of all our
sub-samples is rather similar. One should be, however, aware of still
small statistic of events in the regions at larger $|l|$.

\renewcommand{\arraystretch}{1.0}
\renewcommand{\TableFont}{\scriptsize}
\MakeTableSep{lrcrr}{12.5cm}{Number of microlensing events in the OGLE-II fields}
{
\hline
\noalign{\vskip3pt}
\multicolumn{1}{c}{Field} & $N_{\rm OBS}^{{\mu}{\rm LENS}}$  & $N_{I<19.5}^{\rm STARS}$
&\multicolumn{1}{c}{$N^{{\mu}{\rm LENS}}$ per } \\
& & &\multicolumn{1}{c}{$10^6$ stars}\\
\noalign{\vskip3pt}
\hline
\noalign{\vskip3pt}
BUL$\_$SC1   &   1~~~~~ & 579010 &  1.7~~~~~\\
BUL$\_$SC2   &   4~~~~~ & 650973 &  6.1~~~~~\\
BUL$\_$SC3   &  12~~~~~ & 644230 & 18.6~~~~~\\
BUL$\_$SC4   &  14~~~~~ & 644012 & 21.7~~~~~\\
BUL$\_$SC5   &   5~~~~~ & 280867 & 17.8~~~~~\\
BUL$\_$SC6   &   1~~~~~ & 289813 &  3.5~~~~~\\
BUL$\_$SC7   &   1~~~~~ & 280966 &  3.6~~~~~\\
BUL$\_$SC8   &   0~~~~~ & 205248 &  0.0~~~~~\\
BUL$\_$SC9   &   0~~~~~ & 212660 &  0.0~~~~~\\
BUL$\_$SC10  &   3~~~~~ & 222382 & 13.5~~~~~\\
BUL$\_$SC11  &   0~~~~~ & 199964 &  0.0~~~~~\\
BUL$\_$SC12  &   0~~~~~ & 312079 &  0.0~~~~~\\
BUL$\_$SC13  &   1~~~~~ & 335321 &  3.0~~~~~\\
BUL$\_$SC14  &   7~~~~~ & 414683 & 16.9~~~~~\\
BUL$\_$SC15  &   7~~~~~ & 361358 & 19.4~~~~~\\
BUL$\_$SC16  &   2~~~~~ & 435850 &  4.6~~~~~\\
BUL$\_$SC17  &   2~~~~~ & 454107 &  4.4~~~~~\\
BUL$\_$SC18  &   7~~~~~ & 556327 & 12.6~~~~~\\
BUL$\_$SC19  &   3~~~~~ & 508643 &  5.9~~~~~\\
BUL$\_$SC20  &   6~~~~~ & 692281 &  8.7~~~~~\\
BUL$\_$SC21  &   4~~~~~ & 704858 &  5.7~~~~~\\
BUL$\_$SC22  &   7~~~~~ & 503527 & 13.9~~~~~\\
BUL$\_$SC23  &   6~~~~~ & 455213 & 13.2~~~~~\\
BUL$\_$SC24  &   7~~~~~ & 410873 & 17.0~~~~~\\
BUL$\_$SC25  &   1~~~~~ & 437199 &  2.3~~~~~\\
BUL$\_$SC26  &   5~~~~~ & 446205 & 11.2~~~~~\\
BUL$\_$SC27  &   6~~~~~ & 434129 & 13.8~~~~~\\
BUL$\_$SC28  &   0~~~~~ & 258811 &  0.0~~~~~\\
BUL$\_$SC29  &   2~~~~~ & 257335 &  7.8~~~~~\\
BUL$\_$SC30  &  12~~~~~ & 606406 & 19.8~~~~~\\
BUL$\_$SC31  &   7~~~~~ & 651611 & 10.7~~~~~\\
BUL$\_$SC32  &   3~~~~~ & 680714 &  4.4~~~~~\\
BUL$\_$SC33  &   4~~~~~ & 563023 &  7.1~~~~~\\
BUL$\_$SC34  &   7~~~~~ & 743418 &  9.4~~~~~\\
BUL$\_$SC35  &   5~~~~~ & 614900 &  8.1~~~~~\\
BUL$\_$SC36  &   2~~~~~ & 659023 &  3.0~~~~~\\
BUL$\_$SC37  &   8~~~~~ & 509135 & 15.7~~~~~\\
BUL$\_$SC38  &   5~~~~~ & 572045 &  8.7~~~~~\\
BUL$\_$SC39  &  15~~~~~ & 627658 & 23.9~~~~~\\
BUL$\_$SC40  &   6~~~~~ & 366672 & 16.4~~~~~\\
BUL$\_$SC41  &   8~~~~~ & 394514 & 20.3~~~~~\\
BUL$\_$SC42  &   4~~~~~ & 436339 &  9.2~~~~~\\
BUL$\_$SC43  &   4~~~~~ & 295349 & 13.5~~~~~\\
BUL$\_$SC44  &  10~~~~~ & 183296 & 54.6~~~~~\\
BUL$\_$SC47  &   0~~~~~ & 155703 &  0.0~~~~~\\
BUL$\_$SC48  &   0~~~~~ & 156226 &  0.0~~~~~\\
BUL$\_$SC49  &   0~~~~~ & 148155 &  0.0~~~~~\\
\hline}

Table~3 lists the average number of stars searched for microlensing
events in each of the Galactic bulge fields. In the second column the
number of detected events in each field is provided. Because the number
of searched stars is different by a factor of more than four in our
fields, we normalized the observed number of microlensing events in each
field to one million stars. Normalized number of events is listed in the
last column of Table~3.

\MakeTable{lrrr}{12.5cm}{Average number of microlensing events in the
Galactic bulge}
{
\hline
\noalign{\vskip3pt}
\multicolumn{1}{c}{Line of sight} & \multicolumn{1}{c}{$l$} &
\multicolumn{1}{c}{$b$} &\multicolumn{1}{c}{$N$ per}\\
\multicolumn{1}{c}{BUL$\_$SC} & & & $10^6$ stars \\
\hline
\noalign{\vskip3pt}
5+44	&	$ -0.33$	& $-1.26$ &	36~~~~~\\
3+37	&	$  0.05$	& $-1.83$ &	17~~~~~\\
4+39	&	$  0.48$	& $-2.11$ &	23~~~~~\\
22+23	&	$ -0.38$	& $-3.15$ &	14~~~~~\\
6+7	&	$ -0.20$	& $-5.80$ &	 4~~~~~\\
40+41	&	$ -2.89$	& $-3.20$ &	18~~~~~\\
24+25	&	$ -2.38$	& $-3.46$ &	10~~~~~\\
26+27	&	$ -4.91$	& $-3.51$ &	13~~~~~\\
28+29	&	$ -6.70$	& $-4.52$ &	 4~~~~~\\
47+48+49&	$ -11.21$	& $-2.88$ &	 0~~~~~\\
1+38	&	$  1.02$	& $-3.52$ &	 5~~~~~\\
20+34	&	$  1.52$	& $-2.43$ &	 9~~~~~\\
21+30	&	$  1.87$	& $-2.75$ &	13~~~~~\\
31+32	&	$  2.28$	& $-3.04$ &	 8~~~~~\\
35+36	&	$  3.10$	& $-3.10$ &	 6~~~~~\\
2+33	&	$  2.29$	& $-3.56$ &	 7~~~~~\\
18+19	&	$  4.03$	& $-3.24$ &	 9~~~~~\\
42	&	$  4.48$	& $-3.38$ &	 9~~~~~\\
16+17	&	$  5.19$	& $-3.37$ &	 4~~~~~\\
12+13	&	$  7.85$	& $-3.48$ &	 2~~~~~\\
10+11	&	$  9.69$	& $-3.54$ &	 7~~~~~\\
8+9	&	$ 10.53$	& $-3.88$ &	 0~~~~~\\
14+15	&	$  5.30$	& $ 2.72$ &	18~~~~~\\
43	&	$  0.37$	& $ 2.95$ &	14~~~~~\\
\hline}

Finally, because we typically observe two slightly overlapping driftscan
fields in a given part of the Galactic bulge (see Fig.~1), we averaged
the normalized numbers of events in such adjacent fields to increase
statistic. Table~4 lists our 24 lines of sight in the Galactic bulge
with their Galactic coordinates and the average number of observed
microlensing events per one million stars during seasons 1997--1999 (the
total span of presented observations is about 940 days, \ie 2.58~years).

\begin{figure}[htb]
\psfig{figure=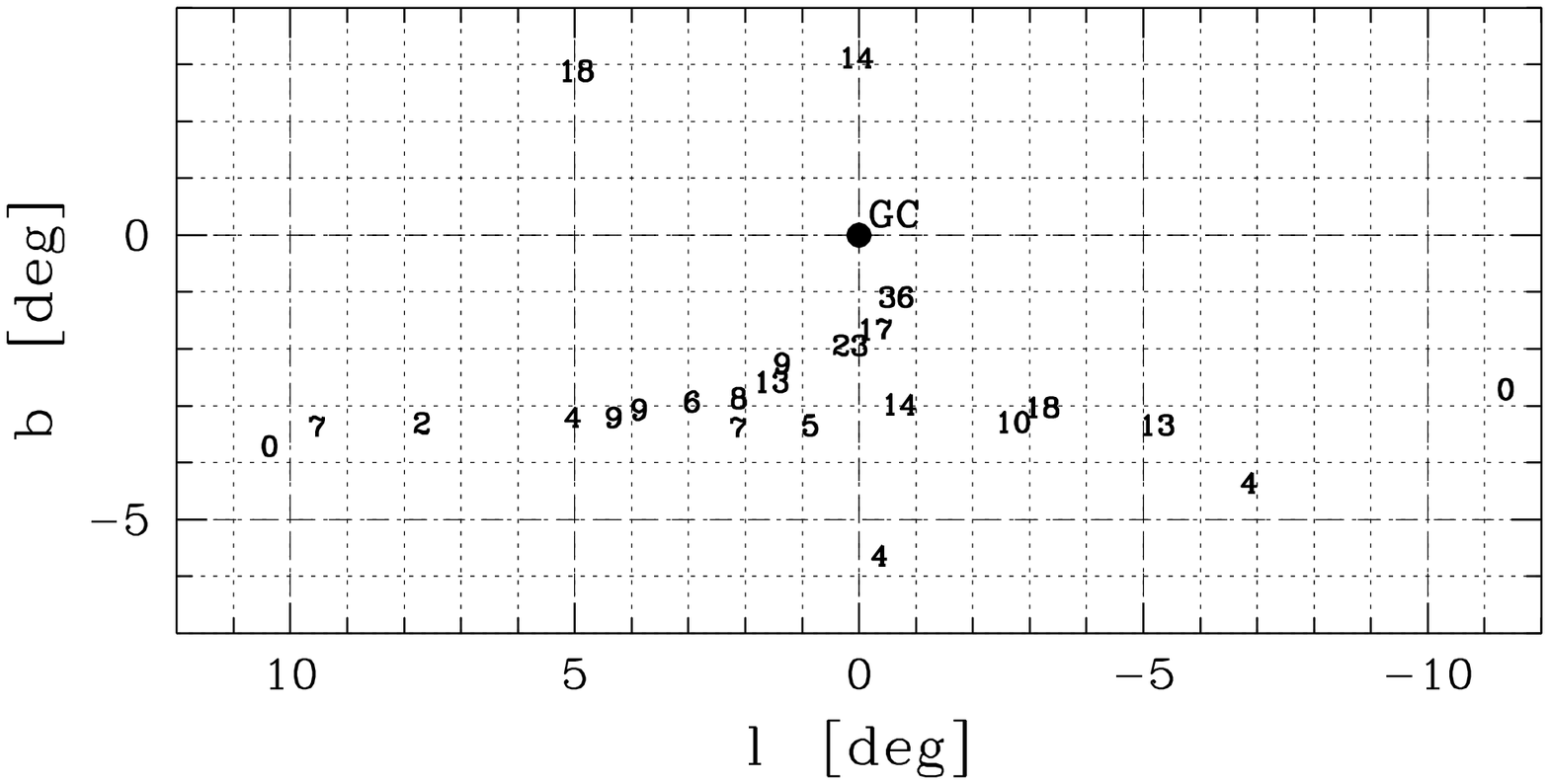,bbllx=15pt,bblly=150pt,bburx=565pt,bbury=430pt,width=12.5cm,clip=}
\FigCap{Rate of microlensing events in the Galactic bulge. The numbers
in the Galactic coordinates grid correspond to the number of
microlensing events per one million stars observed during three bulge
seasons 1997--1999 (2.58~years).}
\end{figure}

Fig.~8 presents the number of events per one million stars observed in
24 lines of sight in the Galactic bulge in years 1997--1999. The number at
given ($l$,$b$) indicates the normalized number of events observed in a
given direction. Of course, one should be aware that the presented
numbers are not the true optical depth. Accurate estimate of the optical
depth would require precise determination of detection efficiency,
assessment  of the blending effect and contribution of the Galactic disk
stars to the stars searched for microlensing events. On the other hand
one can expect that these factors are in the first approximation similar
in so uniformly observed fields. Also the spatial distribution of the
time-scale of events is similar. Therefore the normalized number of
events is likely a crude approximation of the optical depth.

As one can expect the number of events is a strong function of the
Galactic latitude. It falls by almost an order of magnitude when $b$
changes from $b=-1\zdot\arcd3$ to $b=-6\arcd$ (at $l\approx0\arcd$). It
also changes with the Galactic longitude. The numbers of events observed
in the fields at $b\approx-3\zdot\arcd5$ indicate that there is a  clear
dependence on $l$ -- the number of events falls by a factor of 2--4 at
$|l|\approx 10\arcd$ as compared to $l=0\arcd$. There is a noticable
asymmetry with larger number of events at negative $l$. The clear
dependence of the number of microlensing events on $l$ strongly suggests
that the majority of microlensing events are caused by lenses located in
the Galactic bar, inclined to the line of sight toward the Galactic
center,  rather than in the Galactic disk.

At positive $b$  the number of microlensing events in the line of sight
located  at $l=0\zdot\arcd4, b=3\arcd$ is  consistent with that observed
at negative $b$ but one can notice possible excess of events at
$l=5\zdot\arcd3, b=2\zdot\arcd7$.

Looking at the numbers presented in Fig.~8 and Table~4 one should be
aware of the possible bias resulting from somewhat different time
sampling of some fields. While the numbers of observations in most
fields are similar there are six fields with about 40\% larger number of
analyzed frames (see Table~1). These were the fields located closest to
the Galactic center and they were observed with frequency of 3--5
observations per night during a part of the 1997 season for
ultra-short time events. No such events (time scale $t_0<2$~days) were,
however, found. Nevertheless, the number of detected microlensing events
can be larger in more  frequently observed fields resulting in  somewhat
overestimated  ratios between the number of microlensing events close to
the Galactic center and in other directions.

We estimated the possible magnitude of this effect by comparison of the
median time scale, $t_0$, of events from the least-, medium- and
most-frequently observed fields. One could expect that for the most
frequently sampled fields, the efficiency of detection is much higher
for short time scale events and the median $t_0$ of that subsample 
should be significantly shorter than that of the remaining subsamples if
the bias is strong. However, the median $t_0$ is equal to 17.0, 21.5 and
19.7~days for the most-, medium- and least-frequently sampled fields,
respectively. The differences are small indicating that the bias is also
small and can  be in the first approximation neglected. It would be
certainly much stronger if much larger part of events had very short
time scales. 

\begin{figure}[htb]
\psfig{figure=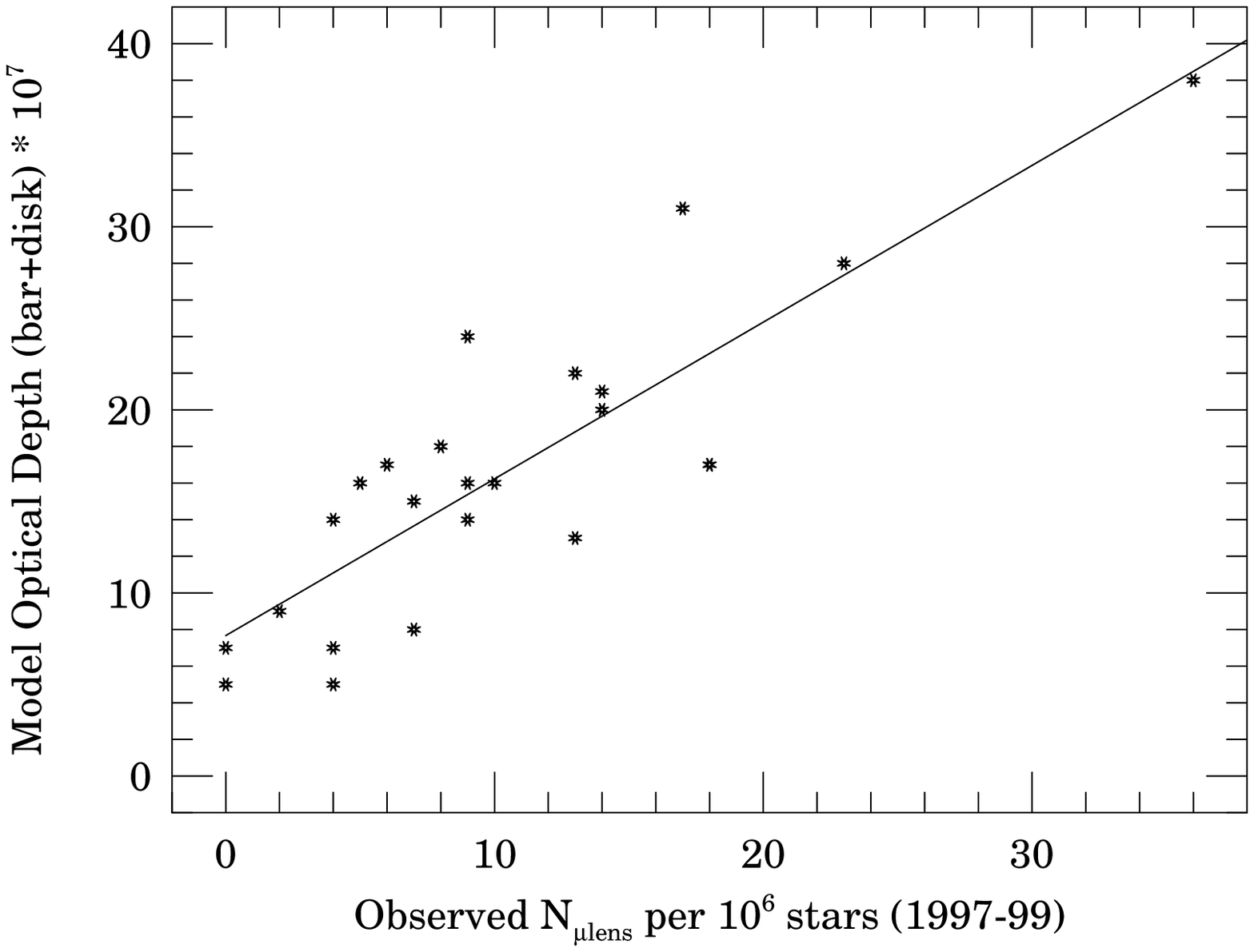,bbllx=60pt,bblly=50pt,bburx=505pt,bbury=405pt,width=12.5cm,clip=}
\FigCap{Correlation between the observed number of microlensing events
and the model optical depth.}
\end{figure}

It is interesting to compare our empirical results with the modeling
predictions. We used the map of the optical depth for the Galactic bar
calculated by Stanek \etal (1997) to derive the optical depth in our
lines of sight. Contribution of the Galactic disk in these directions
was taken from Kiraga (1994) and added to the contribution of the
Galactic bar. Fig.~9 shows the relation between the model optical depth
and the observed number of microlensing events per one million stars in
our lines of sight. The correlation between both values is clearly seen.
The slope of the relation is equal to $0.86\pm0.11$ giving a crude
calibration between the observed numbers and the optical depth. However,
one should also note the potential problem -- the linear relation does
not cross the (0,0) point. This may indicate that the ordinate in Fig.~9
is not that simply related with the optical depth and the observed
normalized number of events is a function not only of  the optical depth
but also of an additional factor like, for instance, spatial dependent
efficiency of detection.  Another possibility is that the model
underestimate significantly the optical depth close to the Galactic
center. In general, however, Figs.~8 and 9 indicate that models of the bar
might provide a reasonable approximation of the Galactic bulge structure
(see also for example Evans 1994, Zhao and Mao 1996, Grenacher \etal
1999)  and that the microlensing will be a very powerful tool in further
constraining the Galactic bulge properties when significantly larger
statistic of events is collected.

While Fig.~8 can provide a first outlook on the distribution of
microlensing events in the Galactic bulge one should be aware that this
is only the first approximation. The statistic of microlensing events,
in particular in the fields with larger $|l|$, is still small.
Observations with the same set-up will be continued during the next
observing season (2000) providing about 60--70 new cases of microlensing
events. After that a large instrumental upgrade to the OGLE-III phase is
planned by implementation of a new mosaic CCD camera. This will increase
the number of discovered events by a factor of 3--5 leading to fast
increase of the statistic of Galactic bulge events. Also larger area of
the Galactic bulge will be monitored. It is also planned to reanalyze
the OGLE-II photometric data with new techniques like image subtraction
method. It would allow to get rid of blending effect uncertainties. By
limiting to well defined population of  the Galactic bulge stars like
for instance red clump giants it will be possible to provide much more
accurate information on the optical depth distribution in the Galactic
bulge in the near future.

The Catalog of Microlensing Events in the Galactic Bulge and all
photometric data presented in this paper are available now to the
astronomical community from the OGLE Internet archive:

\begin{center}
{\it http://www.astrouw.edu.pl/\~{}ogle} \\
{\it ftp://sirius.astrouw.edu.pl/ogle/ogle2/microlensing/gb/}\\
\end{center}
or its US mirror
\begin{center}
{\it http://www.astro.princeton.edu/\~{}ogle}\\
{\it ftp://astro.princeton.edu/ogle/ogle2/microlensing/gb/}\\
\end{center}

\Acknow{We would like to thank Prof.\ Bohdan Paczy\'nski for many
discussions and help at all stages of the OGLE project.  The paper was
partly supported by  the Polish KBN grants 2P03D00814 to A.\ Udalski,
2P03D00916 to M.\ Szyma{\'n}ski, and 2P03D00717 to K.\ \.Zebru\'n.
Partial support for the OGLE  project was provided with the NSF grant
AST-9820314 to B.~Paczy\'nski.}


\begin{references}

\refitem{Albrow, M.D. \etal}{1998}{\ApJ}{509}{687}
\refitem{Albrow, M.D. \etal}{2000a}{\ApJ}{~}{in press (astro-ph/9909325)}
\refitem{Albrow, M.D. \etal}{2000b}{\ApJ}{~}{in press (astro-ph/9910307)}
\refitem{Alcock, C. \etal}{1993}{\it Nature}{365}{621}
\refitem{Alcock, C. \etal}{1995a}{\ApJ}{445}{133}
\refitem{Alcock, C. \etal}{1995b}{\ApJL}{454}{L125}
\refitem{Alcock, C. \etal}{1996a}{\ApJ}{461}{84}
\refitem{Alcock, C. \etal}{1996b}{\ApJL}{463}{L67}
\refitem{Alcock, C. \etal}{1997a}{\ApJ}{479}{119}
\refitem{Alcock, C. \etal}{1997b}{\ApJL}{491}{L11}
\refitem{Alcock, C. \etal}{1999}{\ApJ}{~}{submitted (astro-ph/9907369)}
\refitem{Aubourg, E. \etal}{1993}{\it Nature}{365}{623}
\refitem{Bennett, D.P. \etal}{1999a}{\it Nature}{402}{57}
\refitem{Bennett, D.P. \etal}{1999b}{\it AAS}{195}{3707}
\refitem{Derue, F. \etal}{1999}{\AA}{351}{87}
\refitem{Evans, N.W.}{1994}{\ApJL}{1994}{L31} 
\refitem{Grenacher, L., Jetzer, P., Str\"assle, M, and De Paolis, F.}{1999}{\AA}{351}{775}
\refitem{Kiraga, M.}{1994}{\Acta}{365}{621}
\refitem{Mao, S.}{1999}{\AA}{350}{L19}
\refitem{Mao, S., and Paczy{\'n}ski, B.}{1991}{\ApJL}{374}{L37}
\refitem{Paczy{\'n}ski, B.}{1986}{\ApJ}{304}{1}
\refitem{Paczy{\'n}ski, B.}{1991}{\ApJL}{371}{L63}
\refitem{Paczy{\'n}ski, B.}{1996}{\it ARA\&A}{34}{419}
\refitem{Rhie, S.H., Becker, A.C., Bennett, D.P., Fragile, P.C.,
Johnson, B.R., King, L.J., Peterson, B.A., and Quinn, J.}{1999}{\ApJ}{522}{1037}
\refitem{Stanek, K.Z., Udalski, A., Szyma\'nski, M., Ka{\l}u{\.z}ny, J.,
Kubiak, M., Mateo, M., and Krzemi{\'n}ski, W.}{1997}{\ApJ}{477}{163}
\refitem{Udalski, A., Szyma\'nski, M., Ka{\l}u{\.z}ny, J., Kubiak, M.,
Krzemi{\'n}ski, W., Mateo, M. Preston, G., and Paczy{\'n}ski,
B.}{1993}{\Acta}{43}{289}
\refitem{Udalski, A., Szyma\'nski, M., Ka{\l}u{\.z}ny, J., Kubiak, M.,
Mateo, M. and Krzemi{\'n}ski, W.}{1994a}{\ApJL}{426}{L69}
\refitem{Udalski, A., Szyma\'nski, M., Mao, S., Di Stefano, R., Ka{\l}u{\.z}ny, J., Kubiak, M.,
Mateo, M., and Krzemi{\'n}ski,}{1994b}{\ApJL}{436}{L103}
\refitem{Udalski, A., Szyma\'nski, M., Stanek, K.Z., Ka{\l}u{\.z}ny, J., Kubiak, M.,
Mateo, M. Krzemi{\'n}ski, W., Paczy{\'n}ski, B., and Venkat, R.}{1994c}{\Acta}{44}{165}
\refitem{Udalski, A., Szyma\'nski, M., Ka{\l}u{\.z}ny, J., Kubiak, M.,
Mateo, M. Krzemi{\'n}ski, W., and Paczy{\'n}ski, B.}{1994d}{\Acta}{44}{227}
\refitem{Udalski, A., Kubiak, M., and Szyma\'nski, M.}{1997}{\Acta}{47}{319}
\refitem{Udalski, A., Szyma\'nski, M., Kubiak, M.,Pietrzy\'nski, G., Wo\'zniak, 
P., and {\.Z}ebru\'n, K.}{1998b}{\Acta}{48}{147}
\refitem{Zhao, H.S., and Mao, S.}{1996}{\MNRAS}{283}{1197}
\end{references}
\end{document}